\documentclass[11pt, hidelinks]{article}

\usepackage{amsmath, amssymb, dsfont}

\usepackage[T1]{fontenc}
\usepackage[utf8]{inputenc}
\usepackage{mathptmx}

\usepackage{graphicx,amssymb,verbatim,epsf,amsmath,setspace,float,dsfont, soul, tikz}
\usetikzlibrary{shadows,arrows,trees,shapes,matrix,decorations.pathmorphing,positioning}
\pgfdeclarelayer{background}
\pgfdeclarelayer{foreground}
\pgfsetlayers{background,main,foreground}

\usepackage{graphicx}

\usepackage{cite}
\usepackage{xcolor}

\usepackage{setspace} 
\usepackage{array}
\newcolumntype{C}[1]{>{\centering\let\newline\\\arraybackslash\hspace{0pt}}m{#1}}

\usepackage[labelfont=bf,labelsep=period,justification=raggedright]{caption}

\makeatletter
\renewcommand{\@biblabel}[1]{\quad#1.}
\makeatother

\date{}

\newcommand{\bi}{\begin{itemize}}
\newcommand{\ei}{\end{itemize}}
\newcommand{\bc}{\begin{columns}} 
\newcommand{\ec}{\end{columns}}

\newcommand{\tw}{\textwidth}
\definecolor{dr}{HTML}{CC0000}

\newcommand{\bcat}{$\beta$-catenin }

\numberwithin{equation}{section}
\newcommand{\y}{\checkmark}
\newcommand{\n}{$\times$}

\usepackage{amsthm}

\theoremstyle{definition} 
   
\begin{document}

\title{Mathematical and Statistical Techniques for Systems Medicine: \\The Wnt Signaling Pathway as a Case Study}
\author{
Adam L. MacLean$^{1,2}$, 
Heather A. Harrington$^{1}$, 
Michael P.H. Stumpf$^{2}$,
Helen M. Byrne$^{1,\ast}$
\\
$^{1}$ Mathematical Institute, University of Oxford, Oxford, UK
\\
$^{2}$ Department of Life Sciences, Imperial College London, London, UK
\\
$^{\ast}$E-mail: helen.byrne@maths.ox.ac.uk
}

\maketitle

\onehalfspacing

\section*{Abstract}
The last decade has seen an explosion in models that describe phenomena in systems medicine.
Such models are especially useful for studying signaling pathways, such as the Wnt pathway. 
In this chapter we use the Wnt pathway to showcase current mathematical and statistical techniques that enable modelers to gain insight into (models of) gene regulation, and generate testable predictions.
We introduce a range of modeling frameworks, but focus on ordinary differential equation (ODE) models since they remain the most widely used approach in systems biology and medicine and continue to offer great potential.
We  present methods for the analysis of a single model, comprising applications of standard dynamical systems approaches such as nondimensionalization, steady state, asymptotic and sensitivity analysis, and more recent statistical and algebraic approaches to compare models with data.
We present parameter estimation and model comparison techniques, focusing on Bayesian analysis and coplanarity via algebraic geometry.
Our intention is that this (non exhaustive) review may serve as a useful starting point for the analysis of models in systems medicine. \\

\noindent 
{\bf Keywords:} Wnt signaling, model development, nondimensionalization, asymptotic analysis, parameter inference, algebraic methods, model selection.

\clearpage
\tableofcontents

\section{Introduction}
Despite the growing number of therapeutic options available to clinicians, gaps remain in our fundamental understanding
of many biological processes. Acquiring this additional knowledge requires that we focus on the molecular players that operate in intercellular and intracellular environments. Revealing the complex networks and dynamics that control cellular, tissue- and host-level behavior, may enable us to improve existing treatments and design new drug targets.
\par
Many intercellular signals are initiated by signaling proteins such as cytokines and hormones. When cytokines bind to receptors of a target cell, they trigger a cellular response by signal transduction pathways: multistep sequences of intracellular signaling events and communication between molecules. Most of these molecules are proteins. Enzymes such as kinases and phosphatases, for example, catalyze (respectively) the addition/removal of a phosphate group to/from a substrate, and thus perform a crucial role in relaying information \cite{Alberts02}. Phosphorylation (the addition of a phosphate group) can be associated with protein activation, and information can be communicated downstream, engaging multiple signaling cascades by successive chemical reactions. While some reactions are linear, with the output proportional to the input \cite{Stark:2003cd}, many are complex, involving feedback loops or pathway redundancies. Often the output of these pathways is activation or inhibition of regulatory proteins called transcription factors, which modify gene transcription and the cellular state.
\par
To turn a gene on, an activated transcription factor translocates from the cytoplasm into the nucleus, binds to the enhancer or promoter region of DNA, and RNA polymerase transcribes the DNA template to synthesize RNA. Then messenger RNA (mRNA) leaves the nucleus and enters the cytoplasm where ribosomes translate mRNA into protein \cite{Alberts02}. Conversely, transcription factors may turn a gene off by repressing the recruitment of RNA polymerase. These possible responses thus regulate protein synthesis. In addition to subcellular processes that changes in protein synthesis stimulate, proteins may be released by the cell and act as signaling molecules in other pathways.
\par
Gene regulatory pathways are crucial to the normal functioning of cells, with many diseases caused by dysfunction of one or more pathways. For example, signaling pathways such as NF-$\kappa$B, MAP Kinase, and Wnt/\bcat are involved in a host of cellular processes and functions, including cancer. Due to their complexity, a systems approach is needed to understand normal and aberrant pathway function. Only by building theoretical models that describe how cells signal and validating/updating them using experimental data can we develop new drug therapies that target specific diseases.

The remainder of the chapter is organized as follows. In Section \ref{sect-models}, we review methods used to model signal transduction pathways, and introduce an exemplary enzyme kinetics model. We then describe the biology of Wnt signaling, with reference to relevant models, and  introduce two models of the Wnt signaling pathway that we focus on throughout the chapter to demonstrate various techniques. In Section \ref{sect-a1}, we detail methods that can be used to analyze a particular model and discuss the insight that each approach can generate. In Section \ref{sect-a2}, we introduce techniques that can be used to compare models, including some new methods for systems medicine. We conclude in Section 
\ref{sect-discuss} with a discussion of the different techniques, and ideas for their further application in systems medicine.

\section{Mathematical Modeling}
\label{sect-models}
Signaling pathways are complex and may be difficult to understand by linear logic alone. Theoretical models can be used to gain insight into the dynamics of multiple biochemical interactions. Constructing a mathematical model is a nontrivial task, that requires sufficient understanding of the system to determine not only the type of model that should be used to address a particular question but also the limitations of the model. After reviewing some of the modeling approaches that are used to study signaling pathways, we focus on ordinary differential equation (ODE) models. We introduce basic principles that can be used to construct ODE models and illustrate them by reference to enzyme kinetics and two models of the Wnt pathway. 
 
\subsection{Modeling Approaches for Systems Medicine}
Many processes associated with systems medicine in general, and signaling pathways in particular, can be modeled. These include: gene/protein abundances; gene/protein interactions; abundances of cellular species; the effects of cytokines, chemicals, drugs or other interventions on system or tissue-level phenomena.  Modeling strategies for systems medicine can be classified as either deterministic or stochastic; we describe stochastic approaches briefly here, since the methods introduced in later sections are generally only applicable to deterministic systems.
\par
Deterministic approaches describe systems for which, given full details of the model (parameter values and initial conditions), its time evolution can be determined exactly. This means that if a system is restarted multiple times from the same initial state it will always return to the same future states. Ordinary and partial differential equations (PDEs) are two examples \cite{Murray08}. PDEs with two or more independent variables (e.g., space and time) are more flexible than ODEs, but their simulation and analysis can be computationally expensive. Deterministic methods provide accurate descriptions of population-level behavior if the population sizes are large enough that the effects of random fluctuations can be neglected.
\par
Stochastic approaches describe systems whose temporal evolution has unpredictable elements due to  randomness somewhere in the system. They are popular for modeling biological systems where randomness and heterogeneity abound, and should be used when population sizes are small enough that fluctuations cannot be ignored. In most cases, population averages will be recovered from a stochastic model when the abundances become large enough. One can construct stochastic models of protein dynamics with stochastic differential equations \cite{Gardiner09} (i.e., ODEs with noise terms -- often Gaussian -- added). Such models can be used to study the dynamics of species that fluctuate about a well-defined mean value.  
\par
Stochastic modeling can also be developed via agent-based approaches \cite{Jost05, Gilbert08}. Here, individual agents act according to a set of rules.  For example, within a given pathway, a protein could be phosphorylated or dephosphorylated with probabilities that depend on its environment. Such a framework treats protein species very differently to differential equation methods: each protein is viewed as an autonomous agent and population dynamics emerge in a ``bottom up'' manner.  Whilst such methods may appeal to our intuition about protein heterogeneity, the approach is limited since analyses are often computationally expensive. As such, agent-based models should be used when population-averaged models fail to capture the behavior that the modeler seeks to describe. 
\par
Cellular automata are a subset of agent-based models that impose spatial structure on the system by constraining the agents to lie on a grid, in two or three dimensions \cite{Neumann66, Wolfram:1983wz}. The agents are updated via rules which may be deterministic or stochastic. Each grid point may be occupied by a finite number of cells (typically only one) and the model can accommodate multiple cell types. Cellular automata can account for spatial relationships between different cell types and have the advantage of being easy to interpret biologically. A challenge associated with these models is that the update rules may not translate clearly into biological hypotheses. Additionally, as for other agent-based models, simulation of cellular automata can be computationally expensive. Fitting such models to data is at the limits of what is currently feasible since, despite significant advances in cellular imaging technology, obtaining cell data of sufficient resolution and quality to fit to a model is rare.  
\par
The above overview of modeling approaches is not exhaustive: in limited space, we make no mention of Boolean, semi-quantitative, hybrid, or branching processes. Instead, we continue by explaining how to develop ODE models for signaling pathways.

\subsection{Formulating Mathematical Models of Signaling Pathways}
\label{gen-wnt-models}
In this section our focus is on using ODEs to develop dynamic models of signaling pathways. Two basic principles are integral to the development of such models:
\begin{itemize}
\item {\emph {The Principle of Mass Balance}} states that the rate of change of a species is equal to the difference between the rate at which the species is added to the system and the rate at which it is removed;
\item {\emph {The Law of Mass Action}} states that a reaction proceeds at a rate proportional to the product of its reactants. 
\end{itemize}
If, for example, substrate $A$ is irreversibly phosphorylated by enzyme $B$, to produce $C$ then we write 
\begin{align}
A + B \xrightarrow{r_1} C + B,
\label{rr1}
\end{align}
where $r_1$ is the rate at which phosphorylation occurs. We construct ODEs that describe the dynamics of $A$, $B$ and $C$ by appealing to the Principle of Mass Balance and the Law of Mass Action:   
\begin{align}
\frac{dA}{dt} = -r_1AB, \;\; \frac{dB}{dt} = -r_1 AB + r_1 AB \equiv 0, \;\; \frac{dC}{dt} = r_1AB.
\label{ee1}
\end{align}
By inspecting the above ODEs, it is straightforward to deduce that the following quantities are preserved:
\[ A + C = A_0 + C_0, \quad \mbox{and} \quad B  = B_0, \]
where $A(t=0) = A_0$, $B(t=0) = B_0$ and $C(t=0) = C_0$ are prescribed as initial conditions.
We can exploit these \emph{Conservation Laws} to simplify the governing equations: in this case, we can eliminate both $B$ and $C$ and our model reduces to give
\[ \frac{dA}{dt} = - r_1 B_0 A, \;\; \mbox{with} \;\; A(t=0) = A_0 \;\; \Rightarrow A(t) = A_0 e^{-r_1 B_0 t}.  \]
Thus, substrate levels decay exponentially, at rate $r_1 B_0$.  
 
\subsubsection*{Case Study I: The Enzyme Kinetics Model.}
We now consider a biochemical reaction that is catalyzed by an enzyme. In more detail, the enzyme $E$ binds reversibly with the substrate $S$ to form a complex $C$. While complexed with the substrate, the enzyme converts it into a product $P$ and the enzyme is recovered. We represent these reactions as follows:
\[ E + S \underset{k_{-1}}{\stackrel{k_{1}}{\rightleftharpoons}} C
{\stackrel{k_2}{\rightarrow}}  E + P. \]
By applying the Law of Mass Action to this reaction scheme and appealing to the Principle of Mass Balance, we deduce
that the following system of ODES describe the time-evolution of $S$, $E$, $C$ and $P$:
\begin{align}\label{MM:orig1}
\frac{dS}{dt} &= -k_{1} ES + k_{-1} C, \\
\frac{dE}{dt} &= -k_{1} ES + (k_{-1} + k_2) C, \\
\frac{dC}{dt} &= k_{1} ES - (k_{-1}+k_2)C, \\
\frac{dP}{dt} &= k_2 C. \label{MM:orig4}
\end{align}
If we assume further that $S(t=0) = S_0$, $E(t=0) = E_0$, $C(t=0) = 0$ and $P(t=0) = 0$,
and take suitable combinations of the governing ODEs then we deduce
\[ \frac{d}{dt} (E+C) = 0 \;\; \mbox{and} \;\; \frac{d}{dt} (S+C+P) = 0,\;\; 
\Rightarrow E+C = E_0 \;\; \mbox{and} \;\; S + C + P = S_0, \]
We can exploit these conservation laws to eliminate $E$ and $P$ and obtain the following reduced model:
\begin{align}
\label{MM:reduced1} 
\frac{dS}{dt} &= -k_{1} S(E_0-C) + k_{-1} C, \\
\frac{dC}{dt} & = k_{1} S(E_0-C) - (k_{-1}+k_2)C, \label{MM:reduced2} \\
\mbox{with} \quad S(t=0) &= S_0 \quad \mbox{and} \quad C(t=0)=0. \label{MM:reduced3}
\end{align}

\subsection{Modeling Wnt Signaling}
Wnt signaling is implicated in many biological processes. 
The pathway is activated when Wnt ligands bind to specific receptors on the cell surface, resulting in the stabilization and nuclear accumulation of the transcriptional co-activator $\beta$-catenin. \emph{Canonical} Wnt signaling encompasses cellular responses to external Wnt stimuli mediated by $\beta$-catenin. \emph{Non-canonical} describes cellular signaling and responses to Wnt not mediated by $\beta$-catenin. The canonical Wnt pathway plays a key role in essential cellular processes ranging from proliferation and cell specification during development to adult stem cell maintenance and wound repair \cite{Logan:2004fz}. Dysfunction of Wnt signaling is implicated in many pathological conditions, including degenerative diseases and cancer \cite{Polakis:2000pa, Reya:2005js, Vermeulen:2010ve}. Despite further molecular advances \cite{Goentoro:2009dj, Hernandez:2012cw, Tan:2012hw}, certain details of the dynamics of the pathway are still not well understood. 

\begin{figure}
	\includegraphics[width=\tw]{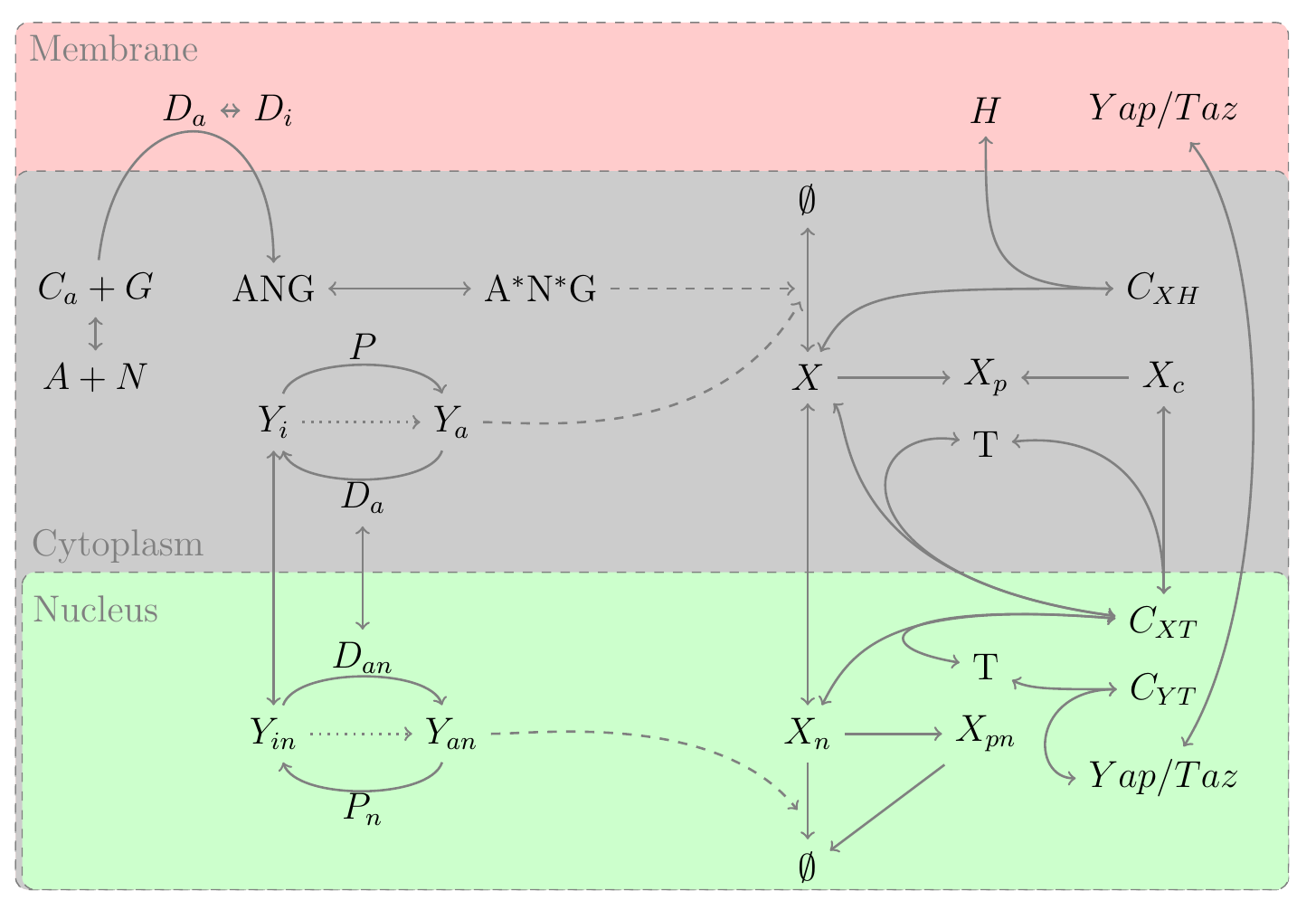}
	\caption{Reaction scheme that incorporates many different Wnt signaling models and additional molecular players (e.g., Yap/Taz). Solid arrows denote direct reactions; long-dashed arrows denote species that act as catalysts in degradation reactions; and dotted arrows denote alternative paths for the direct activation of $Y$. Note that active/inactive forms of $Y$ are equivalent to active/inactive forms of $ANG$.}
	\label{fig-composite}
\end{figure}

The basic steps that constitute canonical Wnt signaling are as follows (although these are not undisputed; discussed below):
Wnt binds to cell-surface receptors Frizzled and LRP5/6 \cite{Reya:2005js} that transduce a signal via a multi-step process involving Dishevelled (Dsh) to the so-called destruction complex (DC). The DC contains forms of Axin, adenomatous polyposis coli (APC), glycogen synthase kinase 3 (GSK-3), and casein kinase 1$\alpha$ (CK1$\alpha$). In the absence of a Wnt signal, the DC actively degrades \bcat -- which is being continually synthesized in the cell -- by binding and phosphorylating the protein and thus marking it for proteasomal degradation. Following Wnt stimulation,  degradation of \bcat is inhibited through phosphorylation of DC member proteins. This leads to accumulation in the cytoplasm of free $\beta$-catenin, which is able to translocate to the nucleus where it can form a complex with T-cell factor (TCF) and lymphoid-enhancing factor (LEF) proteins and, thereby, influence the transcription of target genes associated with processes such as self-renewal and proliferation \cite{LloydLewis:2013ft, Clevers:2012ez}.
\par
In addition to these core mechanisms, evidence for other important processes has been found, some of which may challenge the Wnt signaling paradigm. Spatial localization within the cell has been found to be important not only for \bcat but also for Dsh and DC member proteins including Axin, APC, and GSK-3 \cite{FrancaKoh:2002ia, Wiechens:2004fn, Cong:2004ew, Henderson:2002dl,Itoh:2005kg, Habas:2005dq}. There is also evidence of competitive binding of \bcat to cell membrane proteins such as E-cadherin \cite{Heuberger:2010gl} and intricate cross-talk with the Hippo pathway, this being mediated by Yap and Taz which promote translocation of cytoplasmic \bcat to the nucleus via phosphorylation and then compete with TCF for \bcat in the nucleus \cite{Barry:2013dd} . This spatial organization of Wnt pathway members  may be key to understanding the pathway, as some modeling suggests \cite{Basan:2010kd, MacLean:2015ko}. Equally, an alternative description for the degradation of \bcat exists: in this picture, \bcat can be actively degraded while still bound to the DC, rather than being released marked for degradation \cite{Li:2012jw}. Discriminating between competing hypotheses is needed in order to fully elucidate canonical Wnt signaling: mathematical modeling is a natural framework within which to achieve this.
\par
The first quantitative model of Wnt/\bcat signaling was developed in 2003 \cite{Lee:2003bw}, based on data from {\em Xenopus} extracts. Formulated as a system of ODEs, the model describes known interactions between core components of the canonical pathway, these being Wnt, Dishevelled, GSK3$\beta$, APC, Axin, \bcat and TCF. The DC is assumed to act only in the well-mixed cytoplasm and, hence, only cytoplasmic levels of pathway components are considered. Since its publication, the Lee model has been extended in many ways (for recent reviews of mathematical models of Wnt signaling, see \cite{LloydLewis:2013ft, Kuhl14}). 
The effect of mutations in APC was investigated by Cho {\em et al.} \cite{Cho:2006jt}, the action of Wnt inhibitors was studied by Kogan {\em et al.} \cite{Kogan:2012hl}, and the impact of Wnt-ERK cross-talk  considered by Kim {\em et al.} \cite{Kim:2007ip}. The effect of competition for \bcat with adhesion proteins was investigated by van Leeuwen {\em et al.} \cite{vanLeeuwen:2007ga} while Schmitz showed how shuttling of core proteins between cytoplasm and nucleus could influence pathway dynamics  \cite{Schmitz:2011fe, Schmitz:2013kn}. More recently, a new shuttling model was constructed that accounts not only for exchange of pathway proteins between the nucleus and cytoplasm, but also degradation of \bcat while it is bound to active destruction complex (DC) and activation of the DC by  dephosphorylation of its components \cite{MacLean:2015ko}. Table \ref{tab-models} summarizes the key features of some of these models and Figure 1 illustrates the localization and known interactions between key proteins involved in Wnt signaling.

\begin{table*}[ht]
\begin{center}
\begin{tabular}{| C{5.5cm} |c|c|c|c|c|} \hline
	{\bf Biological Feature } & {\bf Lee } &  {\bf van Leeuwen } & {\bf Schmitz} & {\bf Shuttle} \\\hline 
	\bcat production 
		&\y&\y&\y&\y \\\hline
	\bcat degradation  (independent of DC) 
		&\y&\y&\n&\y \\\hline
	\bcat degradation (dependent on DC) 
		&\y&\y&\y&\y \\\hline
	\bcat sequestration by DC 
		&\y&\y&\y &\y \\\hline 
	\bcat sequestration by APC 
		&\n&\n&\n&\n \\\hline 
	Shuttling of species between cytoplasmic and nuclear compartments 
		&\n&\n&\y&\y \\\hline
	Activation/inactivation of DC 
		&\y &\y &\y &\y \\\hline
	Interaction with adhesion molecules 
		&\n&\y&\n&\n \\\hline
	Two \bcat forms: transcription only and transcription or adhesion 
	 	&\n&\y&\n&\n \\\hline
	DC is represented by its constituent parts 
		&\y &\n &\n&\n \\\hline 
	\bcat binds individual parts APC and Axin as well as DC 
		&\y &\n &\n&\n \\\hline 
	\bcat binds to TCF to promote transcription of target genes  
		&\y&\y&\y&\y \\\hline
\end{tabular}
\caption{Comparison of features across different models of Wnt signaling. For further details see \cite{Lee:2003bw,vanLeeuwen:2007ga,Schmitz:2013kn,MacLean:2015ko}.}
\label{tab-models}
\end{center}
\end{table*}
 
\par
We now present the Lee model \cite{Lee:2003bw} and the Schmitz model \cite{Schmitz:2013kn}, using the notation presented in Table\ \ref{tab-notation}. These models, together with the enzyme kinetics model presented above, will be revisited throughout the chapter to illustrate how the techniques introduced in sections \ref{sect-a1} and \ref{sect-a2} 
are applied to specific models. 
 
\begin{table}
\begin{center}
\begin{tabular}{|c|c|l|}
	\hline {\bf Symbol} & {\bf Species} & {\bf Forms} \\ \hline
	$X$ & \bcat 			
						& $X_{p}$ -- marked for proteasomal degradation  \\ \hline
	$Y$ & Destruction complex		& $Y_a$ -- active 	\\
		& (APC/Axin/GSK3$\beta$)	& $Y_i$ -- inactive 	\\ \hline
	$D$ & Dishevelled 		& $D_a$ -- active \\ 
						& & $D_i$ -- inactive 	\\ \hline
	$A$ & APC			&  	\\
	$N$ & Axin			&  	\\
	$G$ & GSK3$\beta$ 			& 	\\
	$T$ & TCF			& 	\\
	$C$ 	& Complex		& $C_{XY}$ -- complex of X and Y (etc.)	\\ \hline
\end{tabular}
\caption{Definition of notation for the variables used by the Lee and Schmitz models.}
\label{tab-notation}
\end{center}
\end{table}

\subsubsection*{Case Study II: The Lee Model}
\begin{figure}
	\includegraphics[width=\tw]{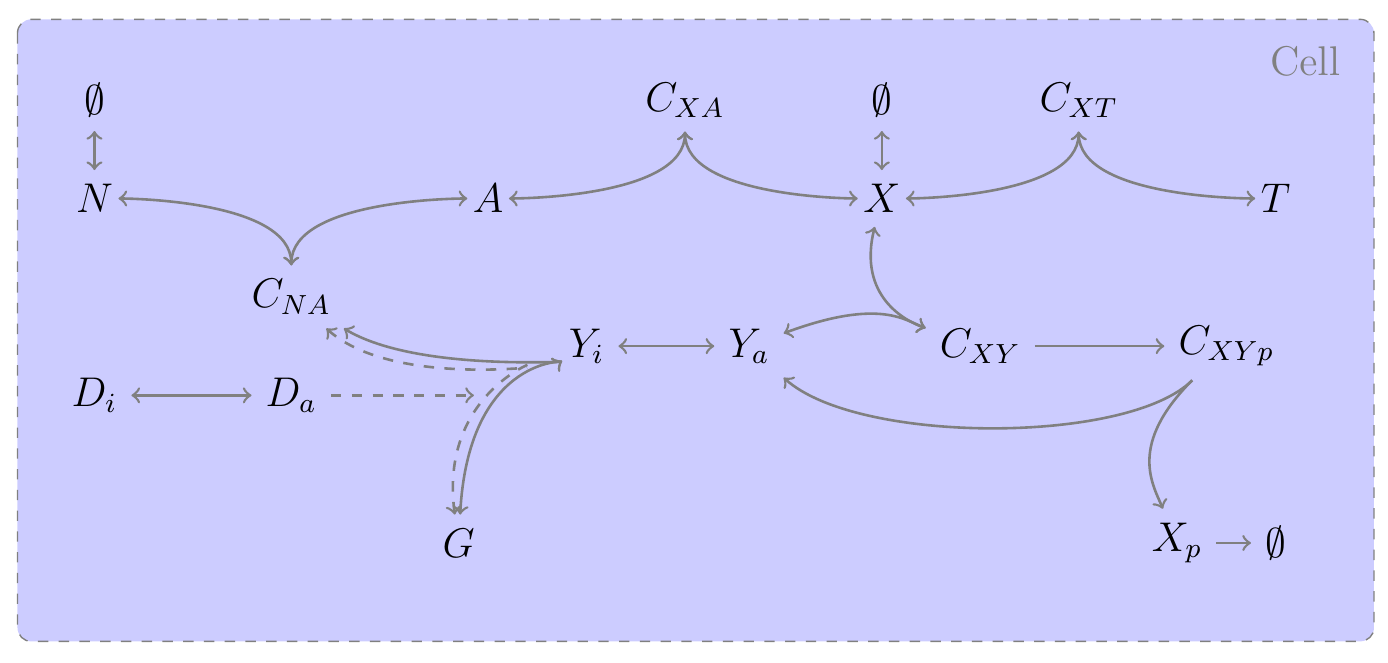}
	\caption{Schematic of the Lee model \cite{Lee:2003bw}, which describes the activation of the destruction complex and its effect on \bcat in a single cellular compartment (cytoplasm and nucleus combined). Notation of the model species is given in Table \ref{tab-notation}. Solid arrows represent reactions and dashed arrows represent catalytic processes.}
	\label{scheme-lee}
\end{figure}

In its original form, the Lee model comprises 15 time-dependent ODEs for protein species and complexes that participate in the Canonical Wnt pathway, the reaction rates being based on mass action kinetics \cite{Lee:2003bw}. The model targets the assembly of the destruction complex from the constituent parts of APC, Axin and GSK3$\beta$. It does not distinguish between nuclear and cytoplasmic compartments, instead assuming that all species are uniformly distributed throughout the cell. A schematic diagram of the reactions described in the Lee model is given in Figure \ref{scheme-lee}. Using the variable names defined in Table \ref{tab-notation} and primes to denote differentiation with respect to time, the ODEs that specify this model are:
\begin{align}
	{D_i}' 	&= -\alpha_1 D_i + \alpha_2 D_a, \label{Lee:orig1} \\
	{D_a}' 	&=  \alpha_1D_i - \alpha_2D_a, \\
	{Y_a}' 	&=  \alpha_3Y_i - \alpha_4Y_a - \alpha_{10}XY_a + \alpha_{11} C_{XY} + \alpha_{13} C_{XYp}, \\
	{Y_i}' 	&= \alpha_6 G C_{NA} - \alpha_5D_aY_i - \alpha_3Y_i + \alpha_4Y_a  - \alpha_7 Y_i, \\
	{G}' 	&= \alpha_5D_aY_i - \alpha_6 G C_{NA} + \alpha_7 Y_i, \\
	{C_{NA}}' 	&= \alpha_5D_aY_i - \alpha_6 G C_{NA} + \alpha_7 Y_i + \alpha_8 N A - \alpha_9 C_{NA}, \\
	{A}'	&= -\alpha_8 N A + \alpha_9 C_{NA} - \alpha_{21} X A + \alpha_{22} C_{XA}, \\
	{C_{XY}}'	&= \alpha_{10} X Y_a - \alpha_{11} C_{XY} - \alpha_{12} C_{XY}, \\
	{C_{XYp}}' 	&= \alpha_{12} C_{XY} - \alpha_{13} C_{XYp}, \\
	{X_p}'	 	&= \alpha_{13} C_{XYp} - \alpha_{14} X_p, \\
	{X}'		&= - \alpha_{10} XY_a  + \alpha_{11} C_{XY} + \alpha_{15} - \alpha_{16} X - \alpha_{19} X T + \alpha_{20} C_{XT} - \alpha_{21} X A + \alpha_{22} C_{XA}, 	\\
	{N}'			&= - \alpha_8 N A + \alpha_9 C_{NA} + \alpha_{17} - \alpha_{18} N, 	\\
	{T}'			&= - \alpha_{19} X T + \alpha_{20} C_{XT} ,	\\
	{C_{XT}}'		&= \alpha_{19} X T - \alpha_{20} C_{XT},	\\
	{C_{XA}}'		&= \alpha_{21} X A - \alpha_{22} C_{XA}. \label{Lee:orig15} 
\end{align}
To facilitate comparison with the Schmitz model (see below), the  non-negative rate constants $\alpha_k$, $k \in (1,2,...,22)$ have been redefined from those used in \cite{Lee:2003bw}. Wnt dependence is incorporated via the parameter $\alpha_1 = \alpha_1(W)$ that controls the activation of Dsh. 
\par
Inspection of Eqs. (\ref{Lee:orig1})-(\ref{Lee:orig15}) reveals that there are four conservations laws:
\begin{align*}
D_0 &= D_i + D_a , \\
G_0 &= G + Y_i + Y_a + C_{XY} + C_{XYp}, \\
A_0 & = A + Y_i + Y_a + C_{XY} + C_{XYp} + C_{XA} + C_{NA}, \\
T_0 &= T + C_{XT}, 
\end{align*}
the constants $D_0, G_0, A_0$ and $T_0$ denoting the (assumed constant) levels of Dishevelled, GSK3$\beta$, APC and TCF initially present in the system. These conservation laws are consistent with experimental 
observations which suggest that levels of these proteins do not fluctuate during Wnt signaling (i.e. they are produced and
degraded at the same rates). They can be used to eliminate 4 variables and, in so doing, to reduce the model from
15 to 11 ODEs. Further simplifications are achieved by assuming that all binding processes, except those for the binding of GSK3$\beta$ to APC/Axin, reach equilibrium rapidly and that all species involving axin are present at low levels. Under these assumptions, and after some algebra, the following expressions for $D_0, G, A, T, X_p, C_{XT}, C_{XYp}$ and $C_{NA}$ 
are obtained:
\[ 
D_i = D_0 - D_a, \;\; G = G_0, \;\; 
A = \frac{A_0}{1+\frac{\alpha_{21}}{\alpha_{22}} X}, \;\; 
T = \frac{T_0}{1+\frac{\alpha_{19}}{\alpha_{20}} X}, \;\; 
X_p = \frac{\alpha_{12}}{\alpha_{14}} C_{XY},  \]
\[ 
C_{XT} = \frac{ X.T_0}{1+\frac{\alpha_{19}}{\alpha_{20}} X}, \;\; 
C_{XA} =\frac{A_ 0 X}{1+\frac{\alpha_{21}}{\alpha_{22}} X},  \;\; 
C_{XYp} = \frac{\alpha_{12}}{\alpha_{13}} C_{XY}, \;\; 
C_{NA} = \frac{\alpha_8}{\alpha_9} \: \frac{A_0 N}{1+\frac{\alpha_{21}}{\alpha_{22}}}, \]
and a reduced system of 7 ODEs for the remaining species is eventually recovered (equations not presented since they are rather involved and less instructive than Eqs. (\ref{Lee:orig1})-(\ref{Lee:orig15})).
In \cite{Lee:2003bw}, and \cite{Kruger:2004kr} this model reduction is performed in an ad-hoc manner; it would be instructive to repeat 
it by first nondimensionalizing the governing equations (see section \ref{nondim}) and using asymptotic analysis to 
perform the model reduction (see section \ref{asympt}). 

\subsubsection*{Case Study III: The Schmitz Model}
\begin{figure}
\begin{center}
	\includegraphics[width=0.7\tw]{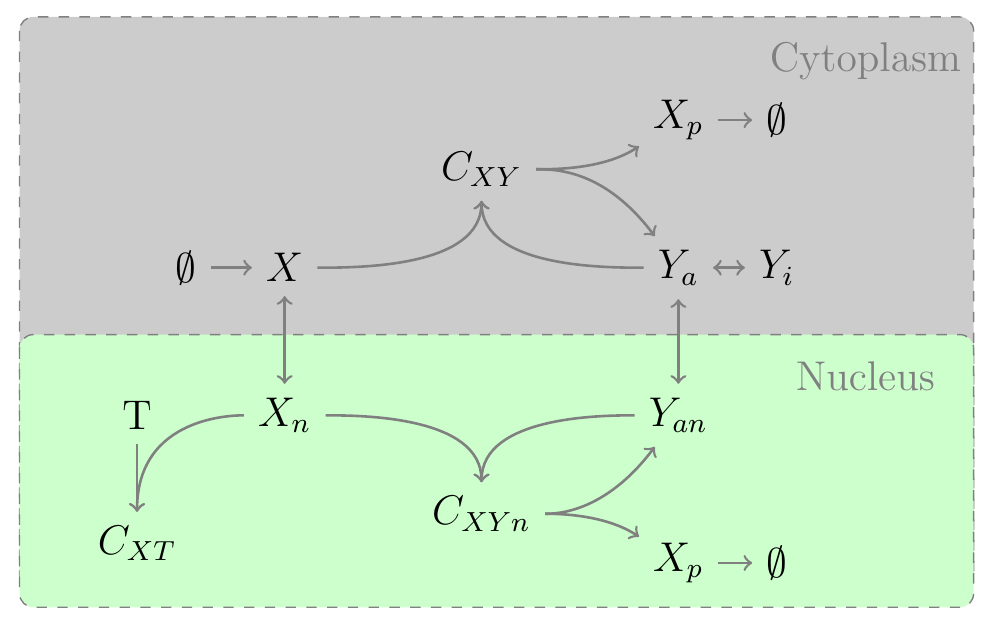}
	\caption{Schematic of the Schmitz model \cite{Schmitz:2013kn}, which describes the interaction between \bcat and the destruction complex in two cellular compartments: cytoplasm and nucleus. Notation of the model species is given in Table \ref{tab-notation}.}
	\label{scheme-sch}
\end{center}
\end{figure}

Like the Lee model, the Schmitz model \cite{Schmitz:2013kn} focuses on the canonical Wnt pathway. Key differences between the Lee and Schmitz models are that the latter distinguishes between the cytoplasm and nucleus and accounts for exchange of \bcat and DC between these compartments (see Table \ref{tab-notation} and Figure \ref{scheme-sch} for further description). 
In each compartment, DC binding to \bcat leads to its phosphorylation, and phosphorylated \bcat is degraded. We use subscript {\em n} to denote species residing in the nucleus with the exception of TCF ($T$) and the $\beta$-catenin-TCF complex ($C_{XT}$); since these species are localized in the nucleus and to facilitate comparison with the Lee model, the subscript is omitted. Using notation that is modified from that used in \cite{Schmitz:2013kn}, the ODEs that define the Schmitz model are:
\begin{align}
\label{Schmitz:orig1}
X' &= \delta_0 + (\delta_2 X_n - \delta_1 X) + (\delta_6 C_{XY} - \delta_5 X Y_a), \\
X_n' &= (\delta_1 X - \delta_2 X_n) + (\delta_9 C_{XYn} - \delta_8 X_n Y_{an})  + (\delta_{12} C_{XT}-\delta_{11} X_n T),\\
X_p' &=\delta_7 C_{XY} - \delta_{13} X_p, \\
X_{pn}' &=\delta_{10} C_{XYn} - \delta_{14} X_{pn},   \\
Y_a' &=(\delta_4 Y_{an} - \delta_3 Y_a)+ (\delta_6 C_{XY} - \delta_5  X Y_a) + \delta_7 C_{XY}  +(\delta_{16} Y_i - \delta_{15} Y_a),  \\
Y_i' &=  \delta_{15} Y_a - \delta_{16} Y_i, \\
Y_{an}' &=(\delta_3 Y_a - \delta_4 Y_{an})  +  (\delta_9 C_{XYn} - \delta_8 X_n Y_{an}) + \delta_{10} C_{XYn}, \\
C_{XY}' &= (\delta_5 X Y_a - \delta_6 C_{XY}) +\delta_7 C_{XY},\\
C_{XYn}' &= (\delta_8 X_n Y_{an} - \delta_9 C_{XYn}) - \delta_{10} C_{XYn}, \\
T' &=\delta_{12} C_{XT} - \delta_{11} X_n T, \\
C_{XT}' & =\delta_{11} X_n T - \delta_{12} C_{XT},  \label{Schmitz:orig11}
\end{align}
where $\delta_k \; (k=1,2,\ldots,17)$ are non-negative rate constants and $\delta_{15} = \delta_{15}(W)$  so that Wnt acts to inactivate the destruction complex in the cytoplasm.
\par
By taking appropriate combinations of Eqs. (\ref{Schmitz:orig1})-(\ref{Schmitz:orig11}), it is straightforward to show that there are two conservation laws:
\begin{equation} 
\label{Schmitz:consn}
Y_i + Y_a + Y_{an} + C_{XY} + C_{XYn} = 
Y_{TOT} \quad \mbox{and} \quad T + C_{XT} = T_0, 
\end{equation}
the constants $Y_0$ and $T_0$ denoting, respectively, the total number of molecules of DC and TCF in the system, as determined from the initial conditions. These identities may be used to reduce the order of the Schmitz model from 11 to 9. As explained below, further systematic simplifications may be possible following model nondimensionalization and parameter estimation.

\section{Techniques for the Analysis of a Specific Model}
\label{sect-a1}
Once model construction is complete, the modeler aims to extract from it new insight. This can be done in a number of ways: if no data are available, standard mathematical techniques can be used to increase understanding of the behavior of the model; however if data are available, then it may be possible to estimate model parameters. 
In this section we describe a number of techniques, some standard and others less so, that can be used to analyze models. We demonstrate these methods by reference to the models of enzyme kinetics and Wnt signaling introduced in Section \ref{sect-models}.

\subsection{Steady State Analysis}
Broadly speaking, the behavior of an ODE model can be categorized as either transient or steady state. The latter describes the behavior at large timescales ($t \rightarrow \infty$). For systems that reach single valued (i.e. not oscillating) steady states, we refer to the long time values that system variables take as the fixed points. Much theory exists for the analysis of fixed points, which can be helpful in characterizing model behavior and predicting the effects of perturbations  \cite{Glendinning94}. We continue by calculating the steady states for the enzyme kinetics model and the Schmitz model (similar analysis can be performed for the Lee model but the resulting expressions are rather involved and therefore omitted).  
 
\subsubsection*{Case Study  I: The Enzyme Kinetics Model (Steady State)}
Setting $\frac{d}{dt}=0$ in Eqs.~(\ref{MM:orig1})--(\ref{MM:orig4}), we deduce that our model for 
enzyme kinetics evolves to the following unique, steady state solution: 
\[ S = 0, \quad E = E_0, \quad C = 0 \quad \mbox{and~} \; P = S_0.  \]
Thus, as expected, the reaction proceeds until all of the substrate $S$ has been converted to product $P$.

\subsubsection*{Case Study III: The Schmitz Model (Steady State)}
Setting $\frac{d}{dt}=0$ in equations (\ref{Schmitz:orig1})--(\ref{Schmitz:orig11}) and manipulating the resulting algebraic equations supplies the following expressions for $Y_{an}, Y_i, X_p, X_{pn}, C_{XY}, C_{XYn}, T$ and $C_{XT}$ in terms of $X, X_n$ and $Y_a$:
\[ 
Y_{an} = \frac{\delta_3}{\delta_4}Y_a, \;
Y_i = \frac{\delta_{15}}{\delta_{16}} Y_a, \; 
X_p = \frac{\delta_7}{\delta_{13}} \: \frac{\delta_5}{\delta_6+\delta_7} XY_a, \;
X_{pn} = \frac{\delta_8}{\delta_{14}} \: \frac{\delta_{10}}{\delta_9+\delta_{10}} XY_a, \]
\[
C_{XY} = \frac{\delta_5}{\delta_6+\delta_7} XY_a, \;
C_{XYn} = \frac{\delta_3}{\delta_4} \: \frac{\delta_8}{\delta_9+\delta_{10}} X_n Y_a, \]
\[ T = \left ( 1 + \frac{\delta_{11}}{\delta_{12}} X_n \right )^{-1} \; T_{TOT}, \;
C_{XT} = \frac{\delta_{11}}{\delta_{12}} \left ( 1 + \frac{\delta_{11}}{\delta_{12}} X_n \right )^{-1} \; X_n T_{TOT},
\]
wherein $Y_a = Y_a(X,X_n)$ satisfies
\[ Y_{TOT} = Y_a \left ( 1 + \frac{\delta_3}{\delta_4} + \frac{\delta_{15}}{\delta_{16}} +
\frac{\delta_5}{\delta_6 + \delta_7} X + \frac{\delta_3}{\delta_4} \: \frac{\delta_8}{\delta_9+\delta_{10}} X_n \right ),\]
while $X_n$  depends linearly on $X$ via
\begin{equation} 
\left ( 1 + \frac{\delta_3}{\delta_4} + \frac{\delta_{15}}{\delta_{16}} \right ) 
= \frac{\delta_5}{\delta_6 + \delta_7}
\left ( \frac{\delta_7}{\delta_{0}} Y_{TOT} - 1\right ) X + \frac{\delta_8}{\delta_9 + \delta_{10}}
 \frac{\delta_3}{\delta_4} \left ( \frac{\delta_{10}}{\delta_{0}} Y_{TOT} - 1 \right ) X_n,  \label{eq:star} 
\end{equation}
and $X$ solves a quadratic of the form
\begin{equation}
0 = {\cal A} X^2 + {\cal B} X + {\cal C}  
\label{eq:diamond} 
 \end{equation}
where the constant coefficients ${\cal A} , {\cal B}$ and ${\cal C} $ are functions of the model parameters. 
For physically realistic solutions, we require $X,X_n>0$.
Therefore, we conclude that this model has at most two steady states and at most one of them may be stable. 

\par
As models increase in complexity, the algebra usually prohibits the construction of analytical expressions for the steady-state solutions. In the following sections we present other methods that can be used to generate insight in such situations.

\subsection{Nondimensionalization}\label{nondim}
 
When a mathematical model is first developed, the independent and dependent variables typically
represent physical quantities (e.g., protein levels) which are measured in dimensional units
(e.g.,  protein levels may be measured as the number of molecules per unit volume or the number of molecules per cell). The model may also contain parameters which relate to physical processes (e.g., reaction rates, Michaelis-Menten constants) and are also dimensional (e.g., rates may be measured per second, per hour or per day). Nondimensionalization involves recasting the model in terms of dimensionless (or unit-less) variables. This process is instructive for several reasons. First, the number of model parameters is typically reduced. Second, the resulting dimensionless parameter groupings can provide useful information about the system's behavior. Further, if estimates of these parameters can be obtained and then compared, it is possible to identify physical processes that dominate on a particular timescale and, thereby, rationale to simplify the governing equations. We illustrate these concepts by nondimensionalizing the enzyme kinetics and Schmitz models.

\subsubsection*{Case Study I: The Enzyme Kinetics Model (Nondimensionalization)}
We introduce the dimensionless variables $\tau, s, e, c$ and $p$  where
\[ t = T \tau, \;\; S = S_0 s, \;\; E = E_0 e, \;\; C = E_0 c, \;\; P = S_0 p. \]
and the timescale $T$ is specified below. It is natural to scale the complex $C$ with $E_0$ since the amount of complex that forms is limited by the amount of enzyme present. If the enzyme is working effectively (i.e., serving as an efficient catalyst), then the amount of product created will be comparable to the amount of substrate. Ttherefore, we scale $P$ with $S_0$ rather than $E_0$. 

There are several possible choices for the timescale $T$. Consider Eq. (\ref{MM:orig1}). Initially, when $C=0$, the maximum rate of uptake of $S$ is $k_1 E_0$ and similarly the initial rate of uptake of $E$ is $k_1 S_0$. The associated timescales are $T_1 = 1/(k_1 E_0)$ and $T_2 = 1/(k_1 S_0)$. Since enzyme levels are typically much smaller than substrate levels (i.e., $E_0 / S_0 = \epsilon \ll 1$), it is clear that $T_2/T_1 =  E_0/S_0 \ll 1$. We conclude that $T_1$ represents a \emph{long} timescale, associated with substrate depletion, while $T_2$ represents a \emph{short} timescale, associated with the initial rapid uptake of enzyme.  

Rescaling on the longer timescale, so that $t = T_1 \tau = \tau/(k_1 E_0)$, Eqs.  (\ref{MM:reduced1})-(\ref{MM:reduced3})  transform to give
\begin{align}
\label{MM:nondim1} 
\frac{ds}{d\tau} &= -s(1-c) + \kappa_e c, \\
\epsilon \frac{dc}{d\tau} & = s(1-c) - \kappa_m c, \label{MM:nondim2} \\
s(\tau=0) &= 1, \quad c(\tau=0) = 0,
\label{MM:nondim4} \\
\mbox{where} \;\; \epsilon = \frac{E_0}{S_0}, \;\; \kappa_e &= \frac{k_{-1}}{k_1 S_0} \;\; \mbox{and} \;\;
\kappa_m = \frac{k_{-1}+k_2}{k_1 S_0}.
\label{MM:nondim3} 
\end{align}
Comparing Eqs. (\ref{MM:reduced1})-(\ref{MM:reduced3}) and (\ref{MM:nondim1})-(\ref{MM:nondim3})
we note that nondimensionalization has reduced the number of model parameters from five to three.
We remark further that in Eq. (\ref{MM:nondim2}), the initial conditions supply $dc(0)/d\tau=1/\epsilon$. Thus, if $\epsilon \ll 1$ then $c$ will initially increase very rapidly on the timescale $\tau$. 
 
\subsubsection*{Case Study III: The Schmitz Model (Nondimensionalization)}
The procedure for nondimensionalizing the Schmitz model is identical to that used for the enzyme kinetics model.
As the dimension of the system increases, and more processes are included, the number of ways to rescale the independent and dependent variables increases rapidly. In such situations, it is important to consider which variables are expected to vary and over what timescale: the answers to these questions should help to identify appropriate scalings.
\par
When studying Wnt signaling, inactivation of the DC plays a key role in the system dynamics and therefore when we nondimensionalize the Schmitz model time is rescaled so that $t = \tau/\delta_{15}$ ($\delta_{15}^{-1}$ is the timescale for inactivation of the DC). Variables relating to free \bcat (i.e $X, X_n, X_p, X_{pn}$) are all rescaled with $\tilde{B} = \delta_0/\delta_{15}$, the amount of \bcat produced during the typical timescale $\tilde{t}$. This scaling eliminates $\delta_0$ from the dimensionless equations (see below). When choosing the scalings for variables involving DC and TCF, we aim to preserve conservation laws. Accordingly, guided by Eqs. (\ref{Schmitz:consn}), we scale   $Y_a, Y_i, Y_{an}, C_{XY}$ and $C_{XYn}$ with $Y_{TOT}$, the total amount of DC in the system. Similarly, we scale $T$ and $C_{XT}$ with $T_{TOT}$, the total amount of TCF in the system. Summarizing, we have
\[ (X, X_n, X_p, X_{pn}) = \tilde{B} (x, x_n, x_p, x_{pn}), \quad
(Y_a, Y_i, Y_{an}, C_{XY}, C_{XYn}) = Y_{TOT} ( y_a, y_i, y_{an}, c_{xy}, c_{xyn}),\]
\[ (T, C_{XT}) = T_0 \: (\theta, c_{x\theta}), \]
where $x(\tau), x_n(\tau), \ldots, c_{x \theta}(\tau)$ are dimensionless variables. Under these scalings, the Schmitz model gives the following nondimensional system:
\begin{align}
\label{Schmitz:nondim1}
x' &=1 + (\tilde{\delta}_2 x_n - \tilde{\delta}_1 x) + (\tilde{\delta}_6 c_{xy} - \tilde{\delta}_5 x y_a), \\
x_n' &= (\tilde{\delta}_1 x - \tilde{\delta}_2 x_n) + (\tilde{\delta}_9 c_{xyn} - \tilde{\delta}_8 x_n y_{an})  + (\tilde{\delta}_{12} c_{x\theta}-\tilde{\delta}_{11} x_n \theta),\\
x_p' &=\tilde{\delta}_7 c_{xy} - \tilde{\delta}_{13} x_p, \\
x_{pn}' &=\tilde{\delta}_{10} C_{xyn} - \tilde{\delta}_{14} x_{pn},   \\
\frac{1}{\omega} y_a' &=\frac{1}{\omega} (\tilde{\delta}_4 y_{an} - \tilde{\delta}_3 y_a)+ (\tilde{\delta}_6 c_{xy} - \tilde{\delta}_5  x y_a) + \tilde{\delta}_7 c_{xy}  +\frac{1}{\omega}(\tilde{\delta}_{16} y_i - y_a),  \\
\frac{1}{\omega} y_i' &=  \frac{1}{\omega} ( y_a - \tilde{\delta}_{16} y_i, \\
\frac{1}{\omega} y_{an}' &=\frac{1}{\omega} (\tilde{\delta}_3 y_a - \tilde{\delta}_4 y_{an})  +  (\tilde{\delta}_9 c_{xyn} - \tilde{\delta}_8 x_n y_{an}) +  \tilde{\delta}_{10} c_{xyn}, \\
\frac{1}{\omega} c_{xy}' &= (\tilde{\delta}_5 x y_a - \tilde{\delta}_6 c_{xy}) + \tilde{\delta}_7 c_{xy},\\
\frac{1}{\omega} c_{xyn}' &= (\tilde{\delta}_8 x_n y_{an} - \tilde{\delta}_9 c_{xyn}) - \tilde{\delta}_{10} c_{xyn}, \\
\frac{1}{\nu} \theta' &= (\tilde{\delta}_{12} c_{XT} - \tilde{\delta}_{11} x_n \theta), \\
\frac{1}{\nu} c_{x\theta}' & = (\tilde{\delta}_{11} x_n \theta - \tilde{\delta}_{12} c_{x \theta} ), \label{Schmitz:nondim11}
\end{align}
where primes denote differentiation with respect to $\tau$ and $\tilde{\delta}_i \; (i=1, \ldots, 16)$ are
the following  dimensionless parameters:
  
\begin{alignat}{5}
&\tilde{\delta}_1 = \frac{\delta_1}{\delta_{15}}, \;
&&\tilde{\delta}_2 = \frac{\delta_2}{\delta_{15}}, \;
&&\tilde{\delta}_3 = \frac{\delta_3}{\delta_{15}}, \;
&&\tilde{\delta}_4 = \frac{\delta_4}{\delta_{15}}, \;
&&\tilde{\delta}_5 = \frac{\delta_5 Y_{TOT}}{\delta_{15}} , \\
&\tilde{\delta}_6 = \frac{\delta_6 Y_{TOT}}{\delta_0}, \;
&&\tilde{\delta}_7 = \frac{\delta_7 Y_{TOT}}{\delta_0}, \;
&&\tilde{\delta}_8 = \frac{\delta_8 Y_{TOT}}{\delta_{15}}, \;
&&\tilde{\delta}_9 = \frac{\delta_9 Y_{TOT}}{\delta_0}, \;
&&\tilde{\delta}_{10} = \frac{\delta_{10} Y_{TOT}}{\delta_0}, \\
&\tilde{\delta}_{11} = \frac{\delta_{11} T_0}{\delta_{15}}, \;
&&\tilde{\delta}_{12} = \frac{\delta_{12} T_0}{\delta_0}, \;
&&\tilde{\delta}_{13} = \frac{\delta_{13}}{\delta_{15}}, \;
&&\tilde{\delta}_{14} = \frac{\delta_{14}}{\delta_{15}}, \;
&&\tilde{\delta}_{16} = \frac{\delta_{16}}{\delta_{15}},\\
& 
&&\omega = \frac{(\delta_0/\delta_{15})}{ Y_{TOT}} \;\; \mbox{and} \;\; 
&& 
&&\nu = \frac{(\delta_0/\delta_{15})}{T_0}.
&&
 \end{alignat}

\subsection{Asymptotic Analysis}\label{asympt}

In applied mathematics, if the (dimensionless) governing equations contain a small parameter, it is common to
assume that there is an asymptotic expansion for the solution, as a power series in the small parameter. As we demonstrate below, this technique can be used systematically to simplify a mathematical model and, in so doing, provide useful information about the dynamics of its components.

\subsubsection*{Case Study I: The Enzyme Kinetics Model (Asymptotics)}
A key assumption of the enzyme kinetics model is that initial enzyme levels are much smaller than substrate levels.
This assumption is represented in the dimensionless model equations via the small parameter
$\epsilon = E_0/S_0 \ll 1$. We exploit this small parameter by seeking a solution to Eqs. (\ref{MM:nondim1})-(\ref{MM:nondim4}) of the form
\begin{equation} 
s(\tau) \sim s_0(\tau) + \epsilon s_1(\tau), \quad c(\tau) \sim c_0(\tau) + \epsilon c_1 (\tau). 
\label{MM:asympt:exp}
\end{equation}
Substituting with Eq. (\ref{MM:asympt:exp}) in the governing equations and equating to zero terms of $O(\epsilon^n)$,
we deduce that, at leading order, $s_0$ and $c_0$ satisfy
\begin{align} 
\frac{ds_0}{d \tau} &= \kappa_e c_0 - s_0 (1-c_0), \\
0 &= s_0 (1-c_0) - \kappa_m c_0, \label{MM:outer2} \\
s_0(0) &= 1, \quad c_0(0) =0. 
\end{align}
Thus the ODE for $c$ reduces to an algebraic relation, giving $c_0$ in terms of $s_0$,  and an ODE for $s_0$,
with the implicit solution
\begin{equation} 
\kappa_m \log s_0(\tau) + s_0(\tau) = A - \kappa \tau, \quad c_0 = \frac{s_0}{\kappa_m + s_0}, \label{MM:outer}
\end{equation}
where $A$ is a constant of integration. A problem arises when we attempt to impose the initial conditions: it is not possible simultaneously to satisfy both initial conditions. This is because the leading order problem is of lower order than the original one. 

In order to resolve this problem, we use \emph{matched asymptotic expansions}. We recall that $c$ varies rapidly 
near $\tau =0$ and, hence, examine the system dynamics near $\tau =0$ by switching to the short timescale $T = \tau/\epsilon$. In terms of $T$, the model becomes
\begin{align} 
\frac{d\tilde{s}}{d T} &= \epsilon (\kappa_e \tilde{c} - \tilde{s} (1-\tilde{c}), \\
\frac{d\tilde{c}}{d T} &= \tilde{s} (1-\tilde{c}) - \kappa_m \tilde{c}, \\
\tilde{s}(0) &= 1, \quad \tilde{c}(0) =0. 
\end{align}
where $\tilde{s}(T) = s (\tau)$, $\tilde{c}(T) = c(\tau)$. As before, we seek asymptotic expansions for $\tilde{s}$ and $\tilde{c}$ in terms of $\epsilon \ll 1$,
of the form specified at Eq. (\ref{MM:asympt:exp}). In this way, we obtain the following leading order solutions for $\tilde{s}_0(T)$ and $\tilde{c}_0(T)$:
\begin{equation} 
\tilde{s}_0(T) = 1, \quad \tilde{c}_0(T) = \frac{1 - e^{-(1+\kappa_m)T}}{1+\kappa_m}. 
\label{MM:inner}
\end{equation}
The above approximate solution is accurate near $\tau = 0$ but not for $\tau = O(1)$, whereas
Eq. (\ref{MM:outer}) is accurate for $\tau=O(1)$ but not for $\tau \ll 1$ . The method of matched asymptotics involves choosing the constant of integration $A$ to match Eqs. (\ref{MM:outer}) and (\ref{MM:inner}) \cite{Kevorkian81}. By imposing the matching conditions
\[ \lim_{\tau \rightarrow 0} (s_0(\tau), c_0(\tau)) = \lim_{T \rightarrow \infty} (\tilde{s}_0(T), \tilde{c}_c(T)), \]
we deduce that $A=1$.
\par 
In practice, similar asymptotic analyses can be used to study ODE models of signaling pathways. As we have seen, such models may involve large numbers of variables and parameters, and estimates for many parameters may be lacking. In such cases, progress can be made by using \emph{order of magnitude} estimates for certain processes. For example, in \cite{Lee:2003bw}, the authors assume that all binding reactions are rapid, apart from the binding of GSK3$\beta$ to APC/Axin. Under this \emph{fast kinetics} assumption, the ODEs for the relevant species reduce to algebraic equations, in the same way that, for the enzyme kinetics model, on the longer timescale the ODE for the complex $c$ reduces to an algebraic relation (see Eq. (\ref{MM:outer2})). 

\par 
To the best of our knowledge, the Schmitz model has yet to be subject to such asymptotic analysis. Referring to 
Eqs. (\ref{Schmitz:nondim1})-(\ref{Schmitz:nondim11}), and by analogy with the asymptotic analysis of the enzyme kinetics model presented above, we note that the dynamics of the system will be strongly influenced by the ratios $\omega$ and $\nu$. 
For example, if typical levels of \bcat are much greater than levels of TCF and DC then we could construct approximate
solutions to the Schmitz model in the limit for which $\nu \ll 1 \ll \omega$. 
Such an analysis of the Lee model was performed by \cite{Mirams:2010ha}. Since the details are rather involved, we summarize the key points below, and refer the interested reader to \cite{Mirams:2010ha} for further details.

\subsubsection*{Case Study II: The Lee Model (Asymptotics)}

\begin{figure}
\begin{center}
	\includegraphics[width=0.7\tw]{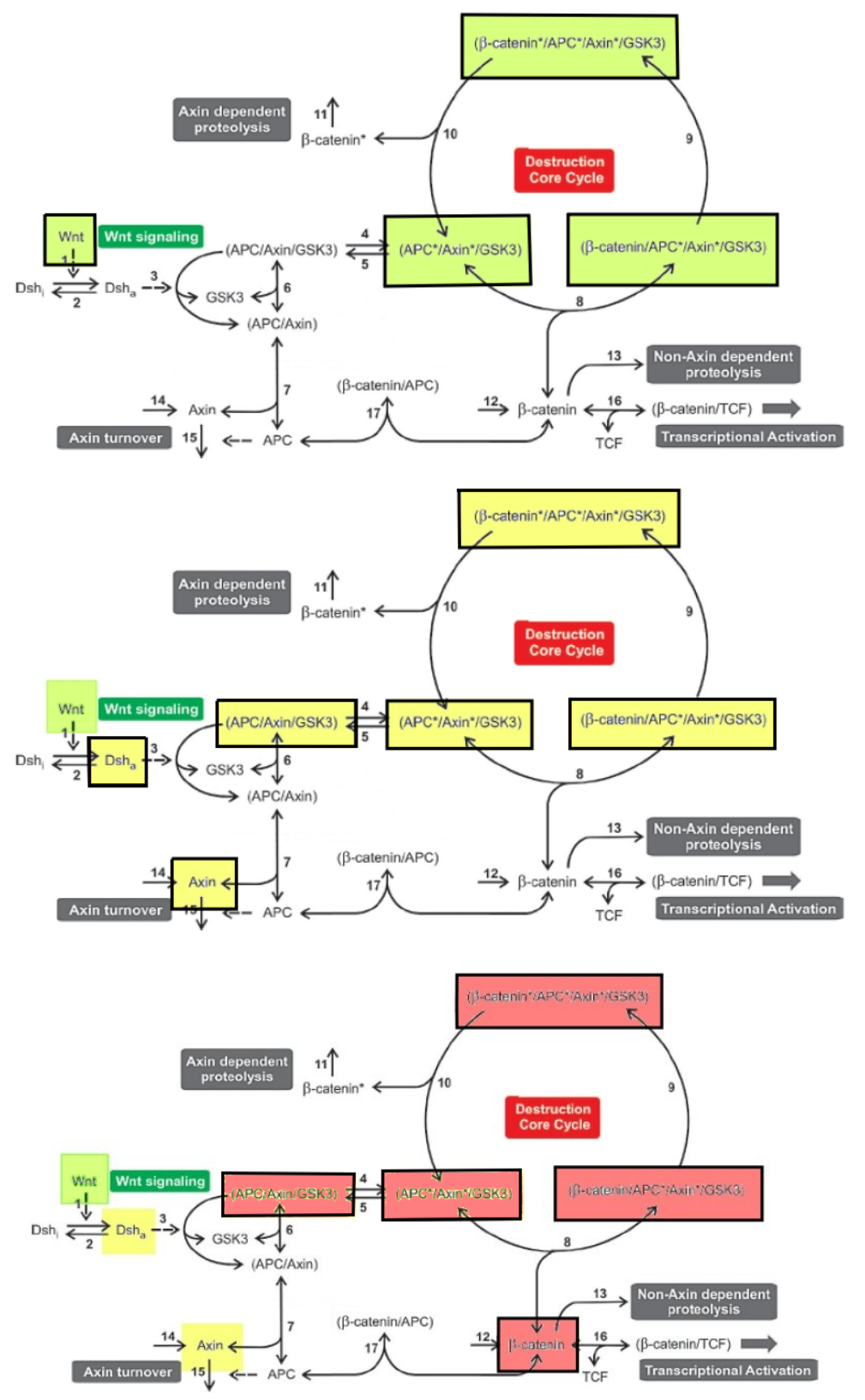}
	\caption{Series of schematics showing which components of the Lee model of Wnt signaling are active on the short (top), medium (middle) and long timescales.  The active components on each timescale are highlighted with bold borders. Figure reproduced from \cite{Mirams:2010ha}, with permission.}
	\label{fig-mirams4}
\end{center}
\end{figure}
 
\begin{figure}
\begin{center}
	\includegraphics[width=0.7\tw]{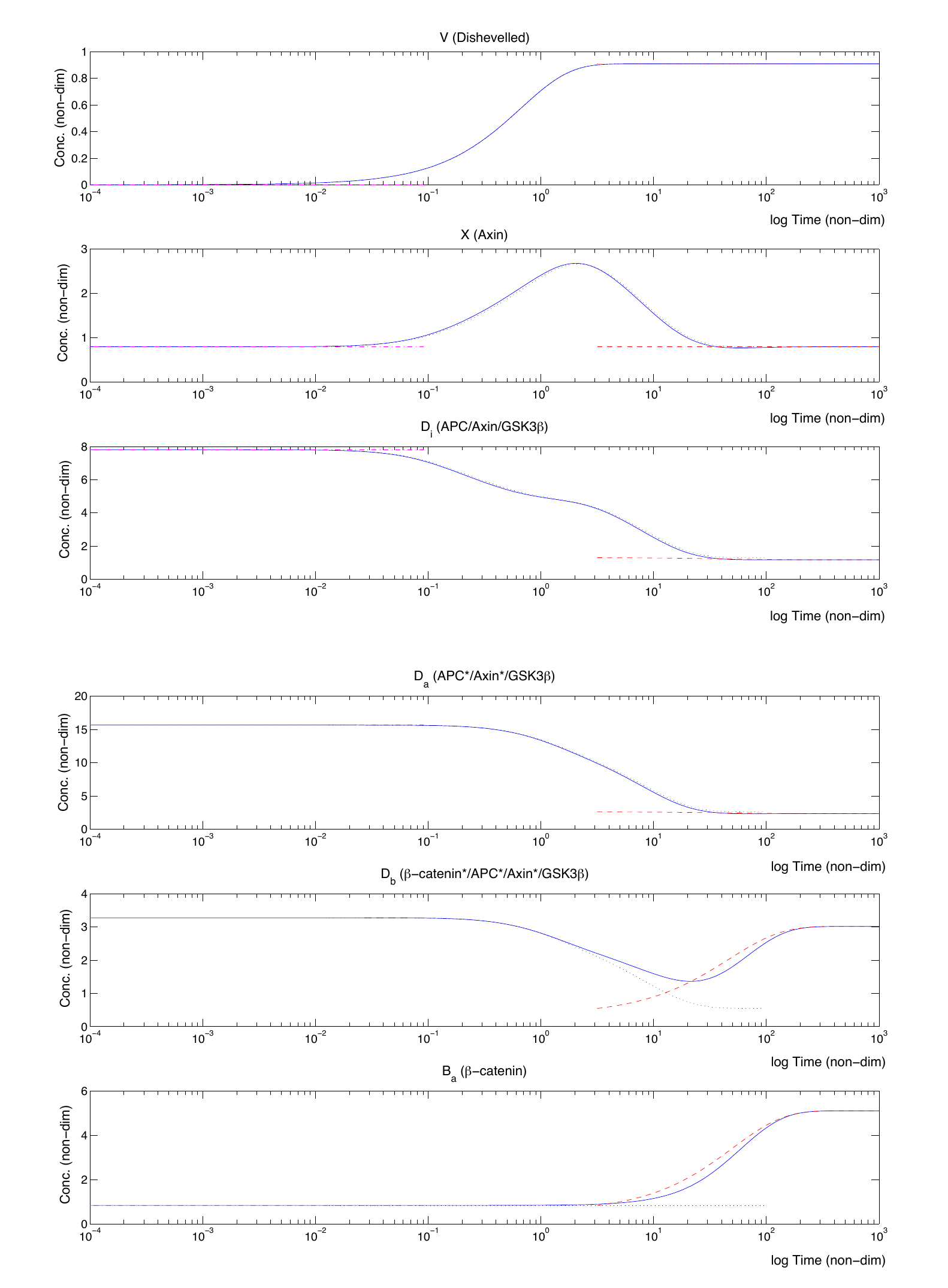}
	\caption{Series of figures showing how the Lee model responds to a Wnt stimulus ($W=1$) that is applied at $t=0$ when the pathway is in equilibrium ($W=0$) at $t=0$. Also shown is the asymptotic solution obtained by matching the short, medium and long time approximations to the Lee model. There is good agreement between the approximate and numerical solutions at all timescales.
Key: numerical simulations of the (dimensionless) Lee model, Eqs. (\ref{Lee:orig1})-(\ref{Lee:orig15}) (solid line); Short, medium and long time approximations are represented by dash-dotted, dotted and dashed lines respectively. Figure reproduced from \cite{Mirams:2010ha}, with permission.}
	\label{fig-mirams5}
\end{center}
\end{figure}

Numerical simulations of the Lee model generated using parameter estimates reported in \cite{Lee:2003bw} (see Figure \ref{fig-mirams5}) suggest that the processes involved in the Wnt signaling pathway act over at least two different timescales. Lee et al.'s parameter estimates indicate that the basal rate at which \bcat is degraded is much smaller than the rate at which the DC becomes inactive. This discrepancy is exploited to define a small parameter,
$\eta = \alpha_{16}/\alpha_{15}$, which is the ratio of the rate at which \bcat undergoes natural decay to the rate at which the DC becomes inactive. The dimensionless parameters are then rescaled by multiplying them by appropriate powers of $\eta$ so that they are $O(1)$. By retaining terms of leading order, the following reduced model is obtained:
\begin{align}
\frac{\textrm{d}D_a}{\textrm{d}t} &= \bar{\alpha}_1W(1-D_a)-\bar{\alpha}_2D_a,\label{Mirams:eq1}\\
\frac{\textrm{d}Y_i}{\textrm{d}t} &= -(\bar{\alpha}_5 D_a+\bar{\alpha}_3 +  \bar{\alpha}_7 )Y_i + Y_a + \frac{\bar{\alpha}_6N}{1+\eta\bar{K}_1 X},\label{Mirams:eq2}\\
\eta\frac{\textrm{d}C_{XY}}{\textrm{d}t} &= \bar{\alpha}_{10} X Y_a - \bar{\alpha}_{11} C_{XY},\label{Mirams:eq3}\\
\frac{\textrm{d}N}{\textrm{d}t} &= \Bigg((\bar{\alpha}_5 D_a+\bar{\alpha}_7)Y_i - \bigg( 
\frac{\bar{\alpha}_6}{(1+\eta\bar{K}_1 X)}+\bar{\alpha}_{18} \bigg  ) N+1\Bigg) \frac{1}{1+\bar{K}_2}, \label{Mirams:eq4}\\
\frac{\textrm{d}Y_a}{\textrm{d}t}&=\frac{\bar{\alpha}_3 Y_i -Y_a - \frac{\textrm{d}C_{XY}}{\textrm{d}t}}{ 1 + \bar{K}_3 X},\label{Mirams:eq5}\\
\frac{1}{\eta}\frac{\textrm{d}X}{\textrm{d}t}&=\bar{\alpha}_{15} - \bar{\alpha}_{10} X Y_a - \bar{\alpha}_{16} X.\label{Mirams:eq6}
\end{align}
We remark that Eq. (\ref{Mirams:eq1}) decouples and if a constant Wnt stimulus is applied ($W(t) = W$, constant) then
\[ D_a \rightarrow \frac{\bar{\alpha}_1 W}{\bar{\alpha}_1 + \bar{\alpha}_2}. \]
We note further that the time derivatives in Eqs. (\ref{Mirams:eq1})-(\ref{Mirams:eq6}) are premultiplied by three different powers of $\eta$. This suggests that model processes act on three distinct timescales, a prediction that  is consistent with the rapid fluctuations and slow increases depicted in Figure \ref{fig-mirams5}.
\par 
As for the enzyme kinetics model (Eq.~\ref{MM:asympt:exp}), it is possible to analyze the reduced Lee model on the short, medium and long timescales for which $t = O(\eta), O(1)$ and $O(\eta^{-1})$ respectively. In each case, asymptotic expansions in powers of the small parameter $\eta$ are sought and used to simplify the governing equations. The results of this analysis can be summarized as follows (see\cite{Mirams:2010ha} for details).
\begin{enumerate}
\item 
Short timescale ($t = O(\eta)$): all model variables except $Y_i$ and $C_{XY}$  are constant, at leading order. The dominant reaction is phosphorylation of \bcat by active destruction complex. 
\item 
Intermediate timescale ($t = O(1)$): the dominant reaction is found to involve inactivation of the destruction complex. 
\item Long timescale ($t = O(\eta^{-1})$): the dynamics are dominated by degradation of free $\beta$-catenin. 
\end{enumerate}
Pathway components acting on the short, intermediate and long timescales are highlighted in Figure \ref{fig-mirams4} while Figure \ref{fig-mirams5} shows good agreement between the approximate solutions and those of the full model.

\subsection{Parameter Analyses}
The selection of model parameters, their physical meaning and numerical value are especially important; therefore, parameter analysis examines the response of the system to changes in parameters. Many methods for estimating parameters depend on time course data. These data generally give a quantitative measure of the variable level, such as mRNA or protein concentration level, at different time points. Testing a model against experimental data is a good way to {\em validate} or {\em invalidate} it; however, gathering experimental data is often too expensive to determine all parameter values and overfitting, i.e., describing noise instead of the relationship, is a risk, as demonstrated for Wnt signaling later in this section. Following parameter estimation (using optimization) or parameter inference (using statistics), a good way to test a model is by performing parameter sensitivity analysis: this evaluates qualitative or quantitative relationships between parameters and their effect on the system outcome \cite{Saltelli:2005cs}.

\subsubsection*{Parameter Estimation and Wnt Data}
Ultimately, every model should be tested against data, a process that can either invalidate the model or provide evidence in its favor, if it provides a good fit under acceptable conditions. 
The aim is to estimate parameters that drive the model close to the data; this can be done using  minimization techniques. Effectively, one calculates an objective function which is defined as the difference between the model simulated for particular value of parameters $\kappa$ and the observations (data), and aims to minimize the error of the objective function, often performed iteratively \cite{Brewer:2008co,Gershenfeld11,BeguerisseDiaz:2012jg}.   
\par
Since publication of the Lee model \cite{Lee:2003bw}, where estimates of the parameters controlling Wnt signaling were based on data from {\em Xenopus} extracts, few studies have quantitatively studied the dynamics of the Wnt pathway. This knowledge gap means that currently it remains difficult to test the models that have arisen in recent years. This problem is not uncommon in systems medicine. We also remark that the {\it{Xenopus}} data gathered by Lee {\em et al.} may be markedly different from those for mammalian Wnt signaling. In \cite{Goentoro:2009dj}, dynamic changes in \bcat levels were investigated in {\em Xenopus} extracts. They demonstrated that absolute levels of \bcat did not dictate the Wnt signaling outcome: rather the \bcat fold-change was the crucial variable. They used the Lee model to test their experimental results and, via sensitivity analysis, identified that the model confirmed their experimental findings.
\par
Quantification of Wnt signaling in mammalian cell lines was undertaken by \cite{Hernandez:2012cw, Tan:2012hw}. Discrepancies with data from {\em Xenopus} extracts (such as higher Axin levels and lower APC levels in mammalian cells) highlight the need for caution in data gathering and for further quantification of the pathway. Since these measurements were made at steady state, they do not yet permit elucidation of transient Wnt signaling. More recent measurements of cytoplasmic and nuclear \bcat in response to a Wnt stimulus provide a valuable first look at the dynamics of the pathway \cite{Tan:2014ii}.
\par 
The above studies provide preliminary insight into the Wnt pathway but much remains to be done. The data are not yet of sufficient quality to discriminate between most models (which typically contain many molecular species). Caution must be taken when applying data. For example, where data generated from non-mammalian systems may be used in a model that addresses clinical outcomes. For systems medicine to have the greatest impact, modeling (with prediction) and experimentation (to test predictions) must proceed iteratively.

\subsubsection*{Parameter Inference}
There are often cases where it is either infeasible or impossible experimentally to determine values for parameters that describe a given model. In such cases, we may be able to estimate (some of) the parameters using statistical inference. In general the aim is to identify the values of the parameters, $\theta$, (ideally including corresponding confidence regions) for which a model best explains the data. 
\par
A reliable way of doing so is to focus on the likelihood $L(\theta)$, which is defined as the probability of observing the data ($x$) given parameters ($\theta$):
\[
L(\theta) := P(x|\theta).
\]
Varying $\theta$ to identify the value for which this probability is maximized gives the maximum-likelihood estimate. There is a rich literature on this topic and how confidence of the estimates can be assessed \cite{Cox79}. 
\par
Likelihood estimates center around the available data. In many circumstances we may have additional information, for example based on biophysical arguments, about which parameter values can be ruled out. Incorporating such {\em prior information} is hard in a pure likelihood framework, but lies at the heart of Bayesian inference \cite{Carlin96}.  Here inferences are based on the {\em posterior distribution} over model parameters. The posterior distribution can be described starting from Bayes rule:
\begin{align}
P(\theta|x) &\propto P(x|\theta)\pi(\theta).
\end{align}
$P(\theta|x)$, the probability of $\theta$ given $x$, is called the posterior probability, $P(x|\theta)$ is the likelihood function, and $\pi(\theta)$ is the prior probability (knowledge about parameters before we begin fitting to data) \cite{Gelman14}. As well as the full (joint) posterior distribution, one may also analyze the marginal posterior distributions which are the individual distributions over each parameter.
\par
In certain cases, such as for large, complex systems, computing the likelihood is impractical. In such cases approximate Bayesian computation (ABC) should be considered \cite{Toni:2009gm}. Instead of the likelihood, a distance function is used to compare the actual data with data simulated by a model, denoted $x_m$. If the underlying model is given by $f = f(x_m|\theta)$, then we express the ABC posterior function by
\begin{align} 
P_{ABC}(\theta|x) \propto  \mathds{1}(\Delta(x,x_m)\le \epsilon) f(x_m|\theta) \pi(\theta)
\end{align}
where $\Delta(a,b)$ denotes a distance measure between $a$ and $b$ and $\epsilon$ is the tolerance level that determines how well real and simulated data should agree.
\par
By evaluating the posterior function, ABC allows the modeler to identify parameter regions that are of interest, and ignore those that are not. Furthermore, the posterior distribution gives information about joint distributions in parameter space and can reveal multivariate dependencies between parameters.

\begin{figure}
\begin{center}
	\includegraphics[width=0.6\tw]{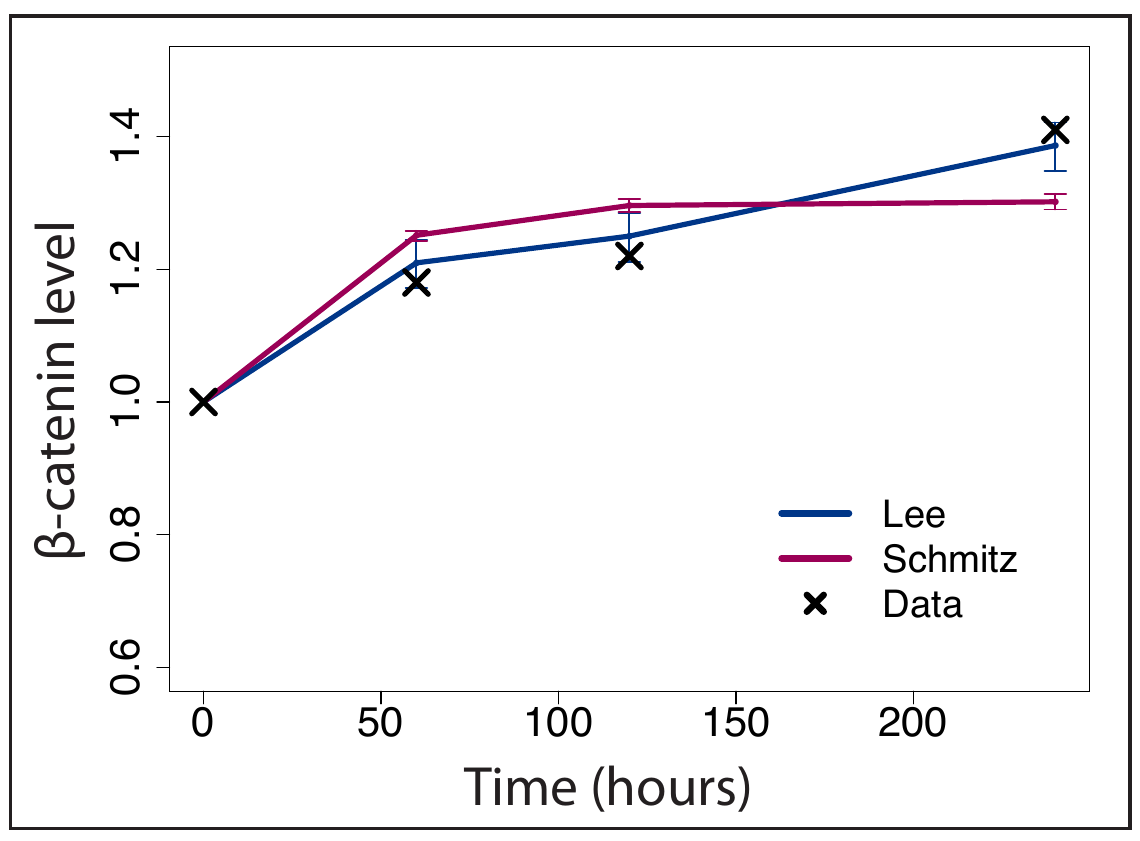}
	\caption{Data published in \cite{Tan:2014ii} were used to fit the Lee and Schmitz models using approximate Bayesian computation for parameter inference. \bcat concentration units were normalized based on their initial values. From the inference, we can see that the Lee model provides a better fit to the data.}
	\label{fig-fits}
\end{center}
\end{figure}

ABC for parameter inference has been implemented in the software package {\tt ABC-SysBio} with support for parallelization \cite{Liepe:2014iw}. For the examples given below, we used the CUDA implementation of {\tt ABC-SysBio} with a Euclidean distance measure between model and data \cite{Liepe:2010eg, Zhou:2011hp}. Proceeding to analyze the Lee and Schmitz models, we do not try to infer all of the model parameters, since this is not possible with the data available, but instead study a 3D subset of parameter space. We choose free parameters that have direct (or strong) influence on the dynamics of $\beta$-catenin, since this is the species for which we have experimental measurements. The data used for fitting are published in \cite{Tan:2014ii}: they describe how the level of \bcat changes over time in the cytoplasm and nucleus, following application of a Wnt stimulus to the system. These data, alongside the results of the parameter inference, are shown in Figure~ \ref{fig-fits}.

\begin{figure}
\begin{center}
	\includegraphics[width=0.8\tw]{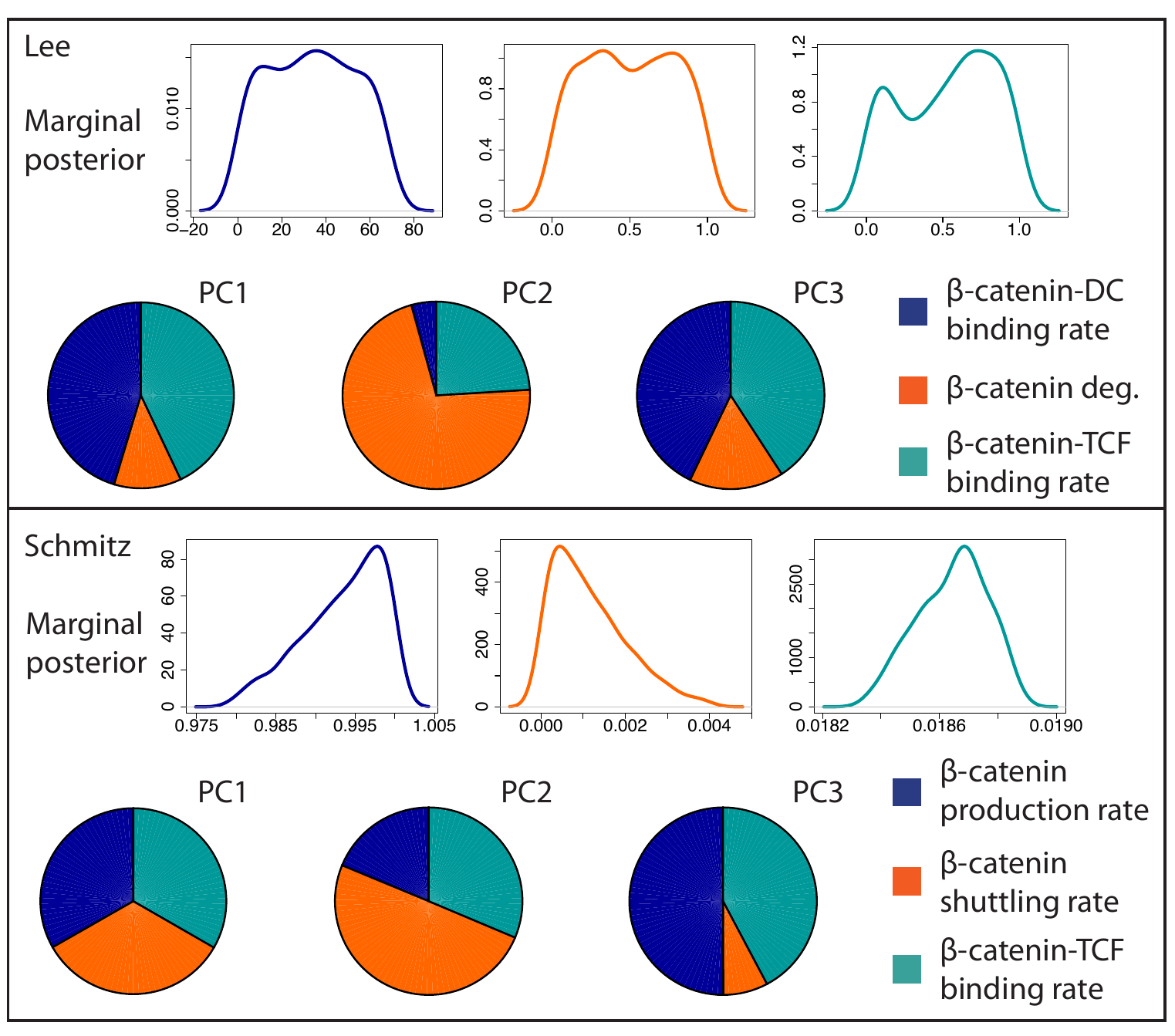}
	\caption{Posterior distributions and sensitivity analysis for the Lee and Schmitz models. Histograms of marginal posteriors for each free parameter in the two models are shown. The marginal posterior is the probability distribution for a single parameter, given data describing \bcat dynamics in cytoplasmic and nuclear compartments \cite{Tan:2014ii}. Principal component (PC) analysis allows us to assess the sensitivity of the parameters to small perturbations: the last PC, PC3, contains the most sensitive parameters. We see that for each model, two parameters dominate PC3 and, thus, are most sensitive in this system.}
	\label{fig-inf-sens}
\end{center}
\end{figure}

For the Lee model, we study the $\beta$-catenin-DC binding rate ($\alpha_{10}$), that has a prior of $[0,100]$, the \bcat degradation rate that is independent of the DC ($\alpha_{16}$), and the binding rate of \bcat to TCF ($\alpha_{19}$). The latter two parameters both have priors of $[0,1]$. The marginal posterior distributions for these three parameters (Figure \ref{fig-inf-sens}) show that the $\beta$-catenin-DC binding parameter takes values over the lower half of its prior range, whereas the other two parameters can take any values spanning the prior range. This suggests that for this model the parameter that has the greatest impact on outcome is the  $\beta$-catenin-DC binding rate, however we note the larger prior range over this parameter.
\par
For the Schmitz model, we study the \bcat production rate ($\delta_0$), the \bcat shuttling rate ($\delta_1$), and the binding rate of \bcat to TCF ($\delta_{11}$). The prior used for each parameter is $[0,1]$ and we see from Figure \ref{fig-inf-sens} that the marginal posterior distributions are relatively stiff: each parameter is constrained to lie within a narrow range relative to its prior. In order to fit the data, the rates of \bcat shuttling and binding to TCF must be low, whilst the rate of \bcat production must be high.

\subsubsection*{Sensitivity Analysis}
Sensitivity analysis investigates how a model responds to perturbations around a set of parameter values and characterizes its robustness: a \emph{robust} system is one for which perturbations of the parameters or initial conditions do not change the outcome. However, many trade-offs between sensitivity and robustness exist \cite{Bluthgen:2003us, Kitano:2006ia, Stelling:2004ed}.
\par
Local sensitivity analysis determines how parameter perturbations affect the output of a system. Estimated or inferred parameters can be used as a baseline for parameter sensitivity. If the output of $dx/dt = f (x,\kappa)$ is approximated by a first-order Taylor series in a neighborhood of reference input values, then the local sensitivity coefficient, $s_{i,j}$ is the partial derivative of the $i^{th}$ state to the $j^{th}$ parameter:
\begin{align}
s_{i,j}(t) &= \frac{\partial x_i(t)}{\partial \kappa_j} ,
\end{align}
The elements $s_{i,j}$ define a sensitivity matrix $S = \partial {\bf x}/\partial {\bf \kappa}$. This local method provides
information about the sensitivity in a given parameter region but not the global sensitivity landscape. Local sensitivity analysis can reveal parameters that are sensitive or robust to perturbations in the region of interest.
\par
Principal component analysis (PCA) offers another way to investigate system sensitivity. This technique can be readily applied to the posterior distribution obtained following Bayesian inference. The principal components are constructed by evaluating the eigenvalues and eigenvectors of the covariance matrix of the parameters: the first principal component (given by the largest eigenvalue) corresponds to the direction in which the posterior is most wide; the last principal component (given by the smallest eigenvalue) corresponds to the direction in which the posterior is most narrow \cite{Secrier:2009ko, Toni:2009gm}. The last few principal components represent the most sensitive (or ``stiff'' parameters) \cite{Gutenkunst:2007gl}.
\par
In Figure \ref{fig-inf-sens}, sensitivity analysis via PCA for the Lee and Schmitz models is shown. The principal components (PC) are ordered 1 -- 3 thus PC3 is the last component and contains the most sensitive parameter combinations. For both models, PC3 is dominated by two parameters: the rates of \bcat binding to the destruction complex (DC) or to TCF for the Lee model ($\alpha_{10}, \alpha_{19}$); and the rates of \bcat production or binding to TCF for the Schmitz model ($\delta_0, \delta_{11}$). These results suggest that the Lee model is more robust to changes in the \bcat degradation rate ($\alpha_{16}$), and that the Schmitz model is more robust to changes in the \bcat shuttling rate ($\delta_{1}$).

\section{Techniques for the Comparison and Discrimination of Models} 
\label{sect-a2}
Given a set of models that describe similar biological phenomena, a challenge is to determine which model best describes the system, given the evidence available. In this section we describe two methods that enable comparison and discrimination between models. The first employs approximate Bayesian computation, introduced above, and has already gained a strong foothold in systems medicine \cite{Liepe:2014iw, MacLean:2014pn, Sunnaker:2013ij, Ratmann:2012db}. The second is model discrimination with the use of algebraic matroids; as far as we know this is a recent addition to the modeler's toolkit and holds great potential for advances in systems medicine.

\subsection{Model Selection via Approximate Bayesian Computation}
Returning now to the Lee and Schmitz models, we consider how to choose between models using ABC model selection. We have already demonstrated how methods for parameter inference, such as ABC, can yield the posterior distributions over the parameters of a model (given data) and discussed briefly how this can be interpreted. For two or more models ($M_i, \; i=1, \ldots,n$) some measure of the evidence for each model is needed \cite{Kirk:2013hq},
\begin{equation}
P(M_i|x) \propto P(x|M_i) \pi(M_i),
\label{eq:Mmodel}
\end{equation}
where (as previously) $x$ represents the data, and $\pi$ the prior probability.
\par
The ABC approach may be extended to parameter inference and model selection simultaneously using a joint space approach \cite{Toni:2009gm}. This may be performed for $M$ models where $M = [M_1, \ldots, M_n]$, by assigning to each model (and parameters therein) a prior distribution and perturbation kernel that designates weights for model transition. The algorithm accepts $N$ particles at the $\epsilon_F$ tolerance, which forms the joint posterior distribution $P(\boldsymbol\alpha,M | \mathbf{\hat{x}})$ and upon marginalizing over parameters, the marginal posterior distribution $P(M | \mathbf{\hat{x}})$ is approximated, providing a measurement for model selection.  Bayesian model selection, like other approaches including the likelihood ratio test or Akaike Information Criteria (AIC), also penalizes over-parameterization.
\par
The AIC for model $M_i$, with $i\in\{1,\ldots,n\}$, is defined as 
 \begin{equation}{
\textrm{AIC}_i=-2\log L(\theta^*_{i};x, M_i)+2k_i,}
\end{equation}
where $L$ is the likelihood, and $\theta_i^*$ and $k_i$ are (respectively) the maximum likelihood parameter and number of parameters in model $M_i$. This criterion, probably the best known model selection tool, makes explicit the penalty for an increased number of parameters. However, as the amount of data increases, the AIC introduces bias and tends to favor models that are over-parameterized. Therefore the Bayesian information criterion (BIC),
 \begin{equation}{
	\textrm{BIC}_i= -2\log L(\theta_i^*;x, M_i) + k_i\log n,}
\end{equation}
may be preferred, as it remains unbiased  for large samples, $n$. The BIC is effectively an approximation to the model probability \eqref{eq:Mmodel}; the penalty term, explicit in the AIC and BIC definitions, is implicit in \eqref{eq:Mmodel}, where it enters via the parameter priors for each model. 
\par
Model selection chooses, from among a set of candidate models, the model that best explains observed data. Two things need to be kept in mind: (i) one model will always be chosen as the best but this does not mean that the model is necessarily a good one; ideally model selection should go hand-in-hand with model checking (and topological sensitivity analysis \cite{Babtie:2014jg}). (ii) model selection depends on the data available for testing the different models; since different data may favor different models, careful experimental design should precede model selection. With these
issues in mind we have the pragmatic choice about which statistical model selection framework to employ. Fully Bayesian, even in an ABC context, is more expensive than identifying the maximum likelihood parameter set and applying AIC or BIC.

\begin{figure}
\begin{center}
	\includegraphics[width=0.6\tw]{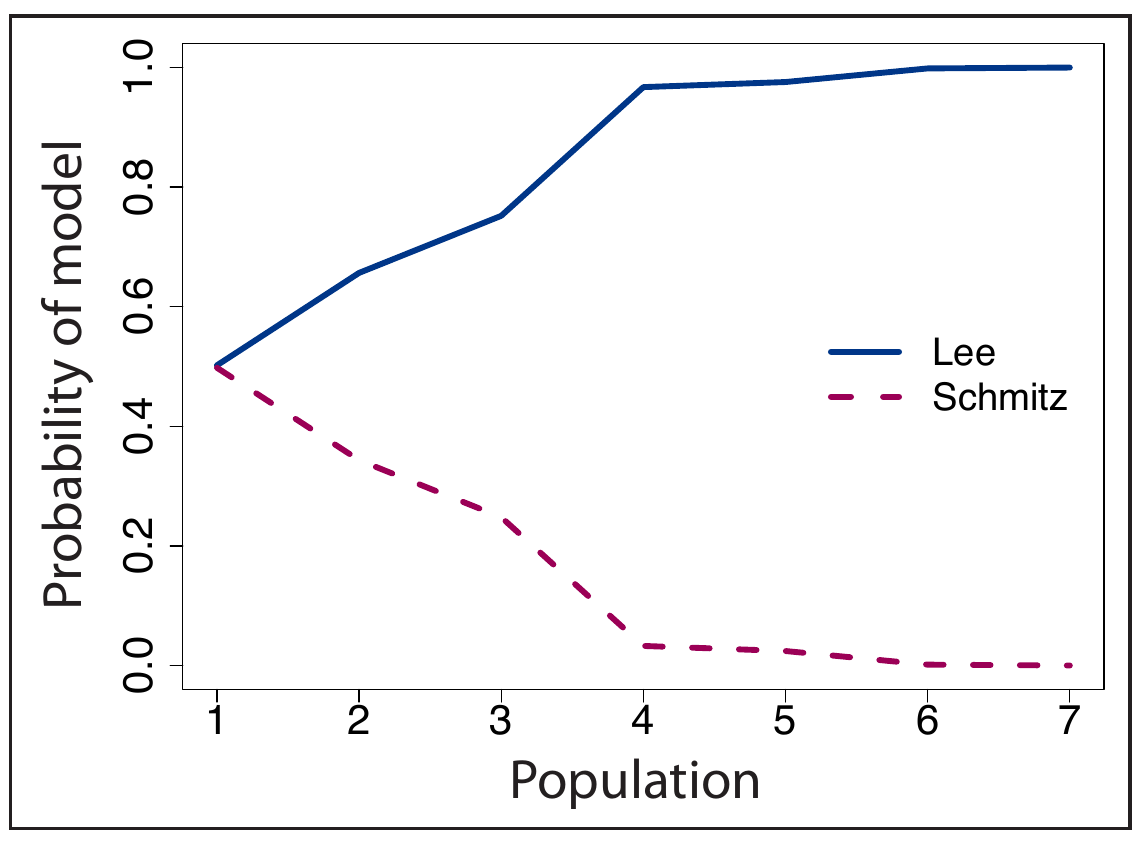}
	\caption{Model selection via ABC for the Lee and Schmitz models. The results show that, over successive populations, evidence in favor of the Lee model grows until there is a high probability that this model will be selected, given the data published in \cite{Tan:2014ii}.}
	\label{fig-model-sel}
\end{center}
\end{figure}

Shown in Figure \ref{fig-model-sel} are the results of ABC model selection for the Lee and Schmitz models, with the probability of the model given for successive iterations (populations). We see that initially both models are equally probable, but subsequently the probability of selecting the Schmitz model drops to close to zero and we conclude that the Lee model is favorable given these data and parameter combinations.

\subsection{Model Discrimination using Parameter-Free (Algebraic) Approaches}
When parameter values are unknown or cannot be estimated from data, one may still be able to discriminate between competing models. We present two approaches, one that requires no data (rather qualitative insight into whether the system can have multiple responses) and another method which requires either highly resolved single cell data or multiple replicates of steady-state measurements.

\subsubsection*{Precluding/Asserting Behaviors via Chemical Reaction Network Theory}
Chemical reaction network theory (CRNT) studies the structure of a model (which can also be described as a network) constructed from chemical reactions without relying on specific parameter values. The aim here is to use such theory to preclude (and sometimes assert) possible qualitative behaviors in the positive orthant, i.e., $\mathbb{R}_{> 0}$. Cases where multiple positive states are stable (i.e., biologically accessible) are of particular biological importance for cellular decision making, for example, differentiation into one of two or more specialized cell lineages.
\par
The field of CRNT initially focused on a structural property of a model called deficiency, which could preclude multiple steady-states \cite{Feinberg:1987tk,Feinberg:1988ei}. Then theorems were proved for precluding/asserting multiple equilibria by studying the cycles in the graph of a network, or the sign of the determinant of the Jacobian; some of these approaches can provide conditions on the parameters for behaviors such as bistability and oscillations \cite{Craciun:2005ub,Craciun:2006et,Craciun:2006ki,Feliu:2011ur,Feliu:2013dz,craciun:pnasu:2006}. An excellent
and comprehensive survey of techniques for multistationarity was written by Joshi and Shiu \cite{Joshi:2014vh}. One main tool for precluding multistationarity of a model is testing whether it is injective (a model, including conservation relations, is {\em injective} if $F(x,\kappa) = F(\tilde{x},\kappa) \Rightarrow x= \tilde{x}$). Here we demonstrate the application of multistationarity tests (developed for chemical reaction networks) to Wnt signaling models.

We begin with the Lee model. First we test injectivity, noting that while injectivity precludes multistationarity, failure of injectivity does not imply multistationarity. We use the algorithms in the CRNT Toolbox to determine whether the system can ever admit multiple positive steady states--multistationarity \cite{CRNT-toolbox}. The Lee model fails injectivity, but cannot admit multiple positive steady states for any values of the system parameters and/or total concentration amounts (algorithms within \cite{CRNT-toolbox}). Conversely, the Schmitz model has the capacity for multiple steady-states; however, as calculated earlier, only one can ever be stable. Therefore, in this example, since both models only can have one stable steady-state,
it is difficult to use only qualitative data to discriminate between them. Clearly, if data suggested two stable states could exist (for example via flow cytometry), and all of the data had the same initial conditions, then one could rule both models out.

\subsubsection*{Model Discrimination using Coplanarity via Algebraic Geometry}
When data from a model clearly supports a specific behavior --- whether monostable, bistable, or oscillatory, qualitative approaches such as those mentioned above may be a good first step for classifying models, especially if the data are not sufficient to estimate parameters. However if steady state data are available, then determining steady-state invariants may be helpful for determining whether a model is compatible with given data using a statistical parameter-free model discrimination method. 

Since often data are not available for all model species, variables must be eliminated. A systematic technique from algebraic geometry proceeds by computing the Gr\"{o}bner Bases of the model variety (studying the model at steady-state) and eliminating unobservable variables. The resulting steady-state invariant enables us to focus on part of the system and to test whether the data suggests that the relationships between species still hold. Notions of dependence and independence between model variables can also be studied using algebraic matroids and were recently applied to steady-state model discrimination \cite{MacLean:2015ko}. 

For smaller models, the steady states can be determined explicitly. For example, for the Schmitz model, 
the steady state values can be expressed in terms of $X$ and $X_n$: all other variables can be eliminated by exploiting conservation laws and using variable substitution (see Eqs.~(\ref{eq:star})-(\ref{eq:diamond})). Either by hand, by computing the matroid, or by using Gr\"{o}bner bases, the polynomial relationship/algebraic dependence between $X$ and $X_n$ in the Schmitz model gives the following invariant:  
\begin{align*}
\mathcal{I} = \delta_0 \delta_3 \delta_4 \delta_6 (\delta_8+\delta_9)X^2 + (\delta_0 \delta_2 \delta_7 \delta_9(\delta_5+\delta_6)- \delta_1 \delta_3 \delta_4 \delta_6(\delta_8+\delta_9))X X_n - \delta_1 \delta_2 \delta_7 \delta_9(\delta_5+\delta_6) X_n^2,
\end{align*}
which vanishes at steady-state (i.e., $\mathcal{I}=0)$. 
Effectively, we aim to test whether the data are coplanar with our model, via the steady-state invariant transformation. 
Model compatibility is determined by computing the coplanarity error ($\Delta$) via the singular value decomposition of the matrix
\[ \left(
\begin{array}{ccc}
	& &  \\
	\hat{X}^2 & \hat{X}_n^2 & \hat{X}\hat{X}_n \\ 
	& & 
\end{array}
\right) \left(
\begin{array}{c}
	\tilde{h}_1 \\
	\tilde{h}_2 \\ 
	\tilde{h}_3 
\end{array}
\right) = 0, \]
where $\hat{X}$ denotes the observed value of species $X$. 
The null hypothesis (that the model is compatible with the data) can be rejected when the coplanarity error (normalized smallest singular value) is less than a statistical bound, which is determined by the Gaussian measurement noise in the data and the invariant structure \cite{Harrington:2012us}. This method was recently applied to $\beta$-catenin localization data (cytoplasmic, $X$; and nuclear, $X_n$) published in \cite{Tan:2014ii, MacLean:2015ko}. The Schmitz model could be ruled out if data were perturbed less than $10^{-5}$ by measurement error/noise; for higher levels of noise, the model is compatible.

\section{Discussion}\label{sect-discuss}
Paradoxically, technological advances can sometimes create new challenges for clinicians. For example, as the number and variety of treatments for cancer increase, it can be difficult to identify the combination of treatments that will most benefit a given patient (if a unique, optimal treatment even exists). The situation is further complicated when we consider the different types of data that can be used as a basis for diagnosis and treatment planning; it is often impossible to integrate the available data by linear thinking alone. Systems medicine aims to address these challenges by developing mathematical and computational tools that integrate different types of information in order to generate objective decisions for patient treatment. In this chapter we have focused on ODE models, a class of models widely used in systems medicine, particularly to study signaling pathways. We have reviewed a variety of techniques that can be used to develop and analyze ODE models, using models of enzyme kinetics and the Wnt signaling pathway as test cases.
\par
Many of the techniques that we have presented are already well-established (such as model development, nondimensionalization, identification of steady state solutions, asymptotic analysis, and parameter sensitivity analysis); however others are less well-known (such as approximate Bayesian computation, chemical reaction network theory, and matroid-informed coplanarity). In addition to the benefit that these methods bring to the field, model development for systems medicine -- in its increasing sophistication -- is helping to stimulate further development and application of mathematical and statistical techniques.
\par 
Many of the challenges in systems medicine arise because most biological processes, including pathways, do not act in isolation. For example, at the subcellular level, pathway cross-talk can have a significant effect on cell function. In particular, there is growing evidence of cross-talk between Wnt and E-cadherin \cite{RamisConde:2008jc},  Wnt and Erk \cite{Kim:2007ip}) and Wnt and the Hippo pathway \cite{Varelas:2010iz}. Even simplistic models of such pathway cross-talk quickly become large and demand sophisticated techniques for their analysis. The situation becomes more complex when we consider the impact of signaling pathways at the multicellular and tissue scales. The impact of Wnt signaling at the multicellular and tissue levels has been studied theoretically, most prominently in models of intestinal crypts \cite{vanLeeuwen:2009ev, Fletcher:2012cc, Murray:2010bo, Murray:2011fu}. These models (for example) introduce spatial dependence by imposing a graded Wnt distribution along the crypt axis \cite{Murray:2010bo} or provide comparison of a continuum model with a cell-based model that incorporates heterogeneity and noise \cite{Murray:2011fu}. In \cite{RamisConde:2008jc}, a multiscale model of interactions between the pathways affecting \bcat and E-cadherin is developed and used to study the role of epithelial-mesenchymal transitions in cancer growth and metastasis, whereas in \cite{Buske:2011es} a simple rule-based model for cross-talk between the Wnt and delta-notch pathways is embedded within discrete epithelial cell agents and used to study cell fate specification within the intestinal crypt. In addition to these theoretical studies (ever growing in complexity), more sophisticated data collection is urgently needed as a basis for hypothesis testing and model (in)validation.
\par
We end by proposing two grand challenges, whose solutions will bear much fruit in systems medicine. The first is to incorporate multiple levels of information --- from biochemical reactions within a single cell, to tissue-level processes --- into cohesive models. The second is to incorporate data which is resolved in space and time into a theoretical framework. There are, of course, many other examples, and work in these areas should provide many exciting opportunities for theoreticians in systems medicine for years to come.

\section*{Acknowledgments}
All authors acknowledge funding from King Abdullah University of Science and Technology (KAUST) KUK-C1-013-04 and the workshop funded by this grant on Model Identification (January 2014). HAH gratefully acknowledges funding from EPSRC Fellowship EP/K041096/1. All authors also thank Gary Mirams for his help with Figures 4 and 5.

\bibliographystyle{spphys}
\bibliography{references}

\end{document}